\documentclass[%
aip,
amsmath,amssymb,
review,%
]{revtex4-1}

\usepackage{graphicx}
\usepackage{dcolumn}
\usepackage{bm}

\usepackage[utf8]{inputenc}
\usepackage[T1]{fontenc}
\usepackage{mathptmx}
\usepackage{etoolbox}

\usepackage{color}
\usepackage{amssymb}
\usepackage{amsmath}
\usepackage{setspace}
\usepackage{graphicx}
\usepackage{subfigure}
\usepackage{booktabs}
\usepackage{makecell}
\usepackage{adjustbox}
\usepackage[ruled,linesnumbered]{algorithm2e}

\makeatletter
\def\@email#1#2{%
	\endgroup
	\patchcmd{\titleblock@produce}
	{\frontmatter@RRAPformat}
	{\frontmatter@RRAPformat{\produce@RRAP{*#1\href{mailto:#2}{#2}}}\frontmatter@RRAPformat}
	{}{}
}%
\makeatother
\begin{document}
	
	\preprint{AIP/123-QED}
	
	\title{Generalized and high-efficiency arbitrary-positioned buffer for smoothed particle hydrodynamics}
	\author{Shuoguo Zhang}
	\author{Yu Fan}
	\affiliation{School of Engineering and Design, 
		Technical University of Munich, Garching, 85748, Germany}
	\author{Yaru Ren}
	\affiliation{School of Engineering and Design, 
		Technical University of Munich, Garching, 85748, Germany}
	\affiliation{State Key Laboratory of Hydraulics and Mountain River Engineering, Sichuan University, Chengdu, Sichuan, China}
	\author{Bin Qian}
	\affiliation{China Southwest Architecture Design and Research Institute Corp. Ltd., Chengdu, Sichuan, China}
	\author{Xiangyu Hu} \thanks{Corresponding author: xiangyu.hu@tum.de}
	\affiliation{School of Engineering and Design, 
		Technical University of Munich, Garching, 85748, Germany}
	
	\date{\today}
	
	\begin{abstract}
		This paper develops an arbitrary-positioned buffer for the smoothed particle hydrodynamics (SPH) method, whose generality and high efficiency are achieved through two techniques.
		First, with the local coordinate system established at each arbitrary-positioned in-/outlet, particle positions in the global coordinate system are transformed into those in it via coordinate transformation. Since one local axis is located perpendicular to the in-/outlet boundary, the position comparison between particles and the threshold line or surface can be simplified to just this coordinate dimension.
		Second, particle candidates subjected to position comparison at one specific in-/outlet are restricted to those within the local cell-linked lists nearby the defined buffer zone, which significantly enhances computational efficiency due to a small portion of particles being checked.
		With this developed buffer, particle generation and deletion at arbitrary-positioned in- and outlets of complex flows can be handled efficiently and straightforwardly.
		We validate the effectiveness and versatility of the developed buffer through 2-D and 3-D non-/orthogonal uni-/bidirectional flows with arbitrary-positioned in- and outlets, driven by either pressure or velocity boundary conditions.   
	\end{abstract}
	
	\maketitle
	
	\section{Introduction}
	\label{Introduction}
	When simulating engineering flows, flow paths typically do not repeat, and in- and outlets may not align to facilitate the flow recirculation \cite{HOLMES2021Novel}. For this kind of smoothed particle hydrodynamics (SPH) flow simulation, the classic periodic boundary condition \cite{Braun2015, Negi2022how}, where newly inflow particles are generated by recycling outflow particles through translation, is incapable for particle generation/deletion at in-/outlet. In comparison, the open boundary condition \cite{FEDERICO2012Simula, Vacondio2012mod, Alvarado-Rodriguez2017, Ferrand2017, Negi2020} can theoretically handle this case easily due to its independent particle handling (i.e., particle generation/deletion) at in-/outlet.
	Several methods have been proposed in SPH literature regarding open boundary conditions. The simplest method is to employ the buffer zone consisting of several layers of buffer particles \cite{Braun2015, Martin2009, Tafuni2018, Shuoguo2022free}. At inlet, when a buffer particle crosses the buffer-fluid interface, a duplicate particle with the same physical state is generated, which then moves freely as an ordinary fluid particle \cite{Wang2019}. Meanwhile, the original buffer particle is recycled to a periodic position at the inlet boundary. At outlet, outflowing particles are deleted once they cross the domain boundary \cite{Lyu2022Toward}.
	A similar procedure, i.e., checking the particle's relative position to the prescribed threshold line (in two dimensions, 2-D) or surface (in three dimensions, 3-D), is also employed in the mirror-axis technique \cite{HIRSCHLER2016Open, KUNZ2016Inflow} for particle generation and deletion. Monteleone et al. \cite{Alessandra2017} utilized a conical scan region with an angle $ \beta $ for particle generation, comparing the angle $ \gamma $ between the cone axis and the line connecting the neighboring particle with the cone vertex to $ \beta $. If there are no effective particles inside the conical scan region, i.e., $ \gamma > \beta/2$ for all neighboring particles, a new particle is created.
	However, when determining whether to establish the conical scan region on a particle, the measured distance $ d_{ip} $ from this particle to the in-/outlet boundary needs to be compared with both the particle spacing $ dp $ and the smoothing length $ h $, i.e., $ dp<d_{ip}<h $.
	In contrast, the semi-analytical boundary treatment \cite{Ferrand2017, LEROY2016new} does not rely on position comparison for particle generation. Specifically, when the evolutive mass of fixed-position inlet vertex particles reaches a certain threshold, a new fluid particle is created at the same location. However, regarding particle deletion, as same to the 
	conical scan approach \cite{Alessandra2017}, the semi-analytical boundary treatment still requires checking whether outflowing particles are outside the domain through position comparison.
	
	Obviously, in these methods mentioned above, comparing the particle position to a prescribed threshold line or surface is the basic and necessary operation for particle generation and deletion.
	In SPH literature, the principles of particle generation and deletion, along with most relevant test cases, are typically illustrated based on simple orthogonal flows, where the in-/outflow direction is parallel to one coordinate axis. 
	As the in-/outlet boundary is generally perpendicular to the in-/outflow direction, 
	the prescribed threshold line or surface for checking the particle's relative position is usually perpendicular to this coordinate axis as well. 
	With this simplification, the position comparison is performed in only one coordinate dimension.
	However, in complex non-orthogonal flows, especially those with multiple arbitrary-positioned in- and outlets, the above solution becomes non-trivial when comparing the particle position to an arbitrary-positioned threshold line or surface. Although directly determining the particle's relative position through the equation of the threshold line or surface can still work, the complex procedure makes it inefficient and non-generalizable.
	In SPH literature, some non-orthogonal flows with multiple arbitrary-positioned in- and outlets have also been conducted \cite{Alessandra2017,LEROY2016new}, but systematic elaborations on how to conduct position comparison at  arbitrary-positioned in-/outlet, especially with a generalized and high-efficiency approach, have rarely been mentioned.
	Furthermore, regarding particle detection, the straightforward way of sequentially checking all fluid particle positions will significantly decrease computational efficiency, especially in 3-D flows with multiple in- and outlets.
	
	In this paper, based on the buffer approach, we introduce the coordinate transformation to develop a generalized and high-efficiency approach for conducting position comparison in particle generation/deletion at arbitrary-positioned in-/outlet.
	The particle positions in the global coordinate system are first transformed into those in the local coordinate system established at in-/outlet. 
	Since one local axis is located perpendicular to the in-/outlet boundary, their position comparison with the threshold line or surface is performed only in this coordinate dimension.
	To further enhance computational efficiency, at one specific in-/outlet, the particle candidates subjected to position comparison are restricted to those within the local cell-linked lists nearby the defined buffer zone.
	The remainder of this paper is organized as follows. 
	First, Section \ref{Riemann-based_WCSPH_method} provides a brief overview of the underlying weakly-compressible SPH (WCSPH) method.
	In Section \ref{Arbitrary_positioned_buffer}, the rationale of arbitrary-positioned uni-/bidirectional in-/outflow buffer is elaborated.
	Section \ref{In_outlet boundary conditions} details pressure and velocity
	boundary conditions implemented at arbitrary-positioned in-/outlet.
	The effectiveness and versatility of the developed arbitrary-positioned buffer are demonstrated by several 2-D and 3-D flow examples in Sections \ref{2D_Numerical_examples} and \ref{3D_Numerical_examples}, including 2-D sloped-positioned VIPO (Velocity Inlet, Pressure Outlet) and PIVO (Pressure Inlet, Velocity Outlet) channel flows, 2-D T- and Y-shaped channel flows, the 2-D rotating sprinkler, the 3-D sloped-positioned pulsatile pipe flow, and the 3-D aorta flow.
	Finally, concluding remarks are given in Section \ref{Conclusion}. 
	The code accompanying this work is implemented in 
	the open-source SPHinXsys library
	\cite{Zhang2021CPC}, 
	and is available at the project website https://www.sphinxsys.org 
	and the corresponding Github repository.
	
	\section{Weakly-compressible SPH method}
	\label{Riemann-based_WCSPH_method}
	\subsection{Governing equations}
	\label{Governing_equations}
	Within the Lagrangian 
	framework, the mass- and momentum-conservation equations 
	are respectively written as 

	\begin{equation} \label{continuity_equation}		
		\frac{d\rho}{dt} = -\rho\nabla\cdot\mathbf{v},
	\end{equation}
	and

	\begin{equation} \label{momentum_conservation_equation}	
		\frac{d\mathbf{v}}{dt} = -\frac{1}{\rho}\nabla p+\nu\nabla^{2}\mathbf{v},
	\end{equation}
	where $ \rho $ is the density, $t $ the time, $ \mathbf{v} $ the velocity, $ p $ 
	the pressure, and $ \nu $ the kinematic viscosity. 
	Under the weakly-compressible 
	assumption, 
	the system of Eqs. (\ref{continuity_equation}) and (\ref{momentum_conservation_equation}) 
	are closed by the artificial isothermal equation of state (EoS) 
	\begin{equation} \label{state_equation}
		p=c_0^2(\rho-\rho_{0}), 
	\end{equation}
	where $ c_{0} $ is the artificial sound speed, 
	and $ \rho_{0} $ the initial reference density. 
	In order to control the density variation around 1$\%$, 
	corresponding to the Mach number $ M\approx 0.1 $ 
	suggested by Monaghan \cite{Monaghan1994Sim}, 
	$ c_{0} $ is set to be $ 10U_{max} $ with $ U_{max} $ denoting the anticipated maximum flow speed. 
	
	\subsection{SPH discretization}
	\label{SPH_discretization}
	Both the mass- and momentum-conservation equations Eqs. (\ref{continuity_equation}) 
	and (\ref{momentum_conservation_equation}) are discretized using the Riemann-based SPH 
	scheme \cite{Zhang2021CPC, Hu2006multi, Zhang2017trans, Zhang2020Dual, ZHANG2021An}, in respect to particle $i$

	\begin{equation} \label{disctretized_continuity_equation}
		\frac{d\rho_{i}}{dt}=2\rho_{i}\sum_{j}\frac{m_{j}}{\rho_{j}}(\mathbf{v}_{i}-
		\mathbf{v}^{*}_{ij})
		\cdot\nabla W_{ij}, 
	\end{equation}
	and

	\begin{equation} \label{disctretized_momentum_conservation_equation}
		\frac{d\mathbf{v}_{i}}{dt}=-2\sum_{j}m_{j}(\dfrac{P^{*}_{ij}}{\rho_{i}\rho_{j}})\nabla 
		W_{ij}+2\sum_{j}m_{j}\frac{\eta\mathbf{v}_{ij} }{\rho_{i}\rho_{j}r_{ij}}\frac{\partial W_{ij}}
		{\partial r_{ij}},
	\end{equation}
	where $ m $ is the particle mass, $ \eta $ the dynamic viscosity, subscript $ j $ the 
	neighbor
	particles, and $\mathbf{v}_{ij}=\mathbf{v}_{i}-\mathbf{v}_{j} $ the relative velocity 
	between 
	particles $ i $ and $ j $. Also, $ \nabla W_{ij} $ denotes the gradient of the kernel 
	function $ W(|\mathbf{r}_{ij}|,h) $, where  $ \mathbf{r}_{ij}=\mathbf{r}_{i}-\mathbf{r}_{j} $, 
	and $ h $ the smooth length. 
	
	Furthermore, $ \mathbf{v}^{*} = U^{*}\mathbf{e}_{ij}+(\mathbf{\overline{v}}_{ij}-\overline{U}
	\mathbf{e}_{ij})$ with $ \mathbf{e}_{ij}=\mathbf{r}_{ij}/r_{ij} $ and 
	$ \mathbf{\overline{v}}_{ij} =(\mathbf{v}_{i}+\mathbf{v}_{j})/2 $ the average velocity 
	between particles $ i $ and $ j $.
	Here, constructed along the unit vector $ \mathbf{-e}_{ij} $ 
	pointing from particle $ i $ to $ j $, the solutions $ U^{*} $ and 
	$ P^{*} $ of the inter-particle one-dimensional Riemann problem can be given 
	as \cite{article2017Chi}

	\begin{equation}
		\begin{cases} \label{Riemann_solver}
			U^{*}=\overline{U}+\dfrac{P_{L}-P_{R}}{2\overline{\rho}c_{0}}\\[3mm]
			P^{*}=\overline{P}+\frac{1}{2}\overline{\rho}c_{0}(U_{L}-U_{R})\\[3mm]
		\end{cases},
	\end{equation}	
	with the initial states of the Riemann problem
	\begin{equation} 
		\begin{cases} \label{initial_states}			
			(\rho_{L},U_{L},P_{L})=(\rho_{i},\mathbf{-v}_{i}\cdot \mathbf{e}_{ij},p_{i})\\[3mm]
			(\rho_{R},U_{R},P_{R})=(\rho_{j},\mathbf{-v}_{j}\cdot \mathbf{e}_{ij},p_{j})
		\end{cases}.
	\end{equation}	
	Here, $ \overline{U}=(U_{L}+U_{R})/2 $, $\overline{P}=(P_{L}+P_{R})/2 $, 
	and $\overline{\rho}=(\rho_{L}+\rho_{R})/2 $ are inter-particle averages. $ L $ and $ R $ 
	denotes the initial left and right states, respectively.
	
	However, the utilization of the original intermediate pressure $P^{*}$ in 
	Eq. (\ref{Riemann_solver}) during the SPH discretization may lead to an excessive 
	dissipation. To address this issue, a supplementary low dissipation Riemann 
	solver \cite{article2017Chi} is employed, which incorporates a modification on 
	$P^{*}$ while maintaining the intermediate velocity $ U^{*} $ in Eq. (\ref{Riemann_solver}) 
	unconstrained
	\begin{equation} \label{low_dissipation_limiter}
		P^{*}=\overline{P}+\frac{1}{2}\beta\overline{\rho}(U_{L}-U_{R}),
	\end{equation}   
	where $ \beta = min\:(3max(U_{L}-U_{R},0),c_{0}) $ represents the low dissipation limiter.
	
	In order to increase the computational efficiency, 
	the dual-criteria time stepping method \cite{Zhang2020Dual} is employed herein.
	In addition, to decrease the accumulated density error during long-term simulations, 
	the particle density is re-initialized before being updated 
	in continuity equation Eq. (\ref{disctretized_continuity_equation}) 
	at each advection time step with \cite{Zhang2021CPC}
	\begin{equation} \label{density_summation}
		\rho_{i}=\rho_{0}\dfrac{\sum W_{ij}}{\sum W^{0}_{ij}},
	\end{equation}
	where the superscript $ 0 $ represents the reference value in the initial configuration.
	\subsection{Transport-velocity formulation}
	\label{Transport_velocity_formulation}
	To avoid particle clumping and void regions in SPH simulations where the tensile instability 
	is present \cite{Lind2012Incom, OGER201676SPH}, regularization is implemented 
	to maintain a uniform particle distribution \cite{litvinov2015towards, Negi2022Tec}.
	Currently, the particle shifting technique (PST) 
	\cite{Lind2012Incom, OGER201676SPH, Negi2022Tec, Skillen2013, KHAYYER2017236Comp, SUN201725} 
	and the transport-velocity formulation 
	(TVF) \cite{Zhang2017trans, Adami2013trans} are two typical schemes to address this issue. 
	In the present work, the TVF scheme is adopted, and the particle advection velocity 
	$ \widetilde{\mathbf{v}} $ is rewritten as

	\begin{equation} \label{transport_velocity}
		\widetilde{\mathbf{v}}_{i}(t + \delta t)=\mathbf{v}_{i}(t)+
		\delta t\left(\frac{\widetilde{d}\mathbf{v}_{i}}{dt}-p_{max}\sum_{j} \frac{2m_{j}}
		{\rho_{i}\rho_{j}} \frac{\partial W_{ij}}{\partial r_{ij}}\mathbf{e_{ij}} \right),
	\end{equation}
	where the global background pressure $p_{max}=\lambda\rho_{0}\mathbf{v}_{max}^{2}$ 
	with $ \mathbf{v}_{max} $ denoting the maximum particle velocity 
	at each time step \cite{Zhang2020Dual}. The empirical coefficient $ \lambda $ is chosen within the range of 5.0 to 10.0 \cite{Zhang2017trans}. 
	
	Since the TVF scheme is applicable only to inner fluid particles far away from in-/outlet 
	boundaries \cite{Adami2013trans}, accurately identifying boundary particles 
	is a necessary prior. To avoid misidentifying and missing 
	boundary particles \cite{Shuoguo2022free, Dilts, Haque, Lee2008}, 
	the spatio-temporal identification approach \cite{Shuoguo2022free} is utilized in 
	present work, where approximately three layers of fluid particles are identified 
	as boundary particles.    
	
	\section{The rationale of arbitrary-positioned buffer}
	\label{Arbitrary_positioned_buffer}	
	In this section, we introduce the rationale of arbitrary-positioned buffer working for uni- and bidirectional flows, respectively. Although the schematic provided here is based on channel flow, this arbitrary-positioned buffer can also be implemented in free-surface open-channel flows with arbitrary-positioned in- and outlets. In this work, the C2 Wendland kernel
	\cite{Wendland1995} with a compact support of $2h$ is employed.
	Since the present smoothing length $ h = 1.3dp $ following the general practice,
	the buffer zone should consist of at 
	least three layers of particles to ensure full support for 
	the inner-domain particles next to the buffer.
	\subsection{Unidirectional in-/outflow buffer}
	\label{unidirectional_in_outflow}	

	\begin{figure*}
		\includegraphics[width=\textwidth]{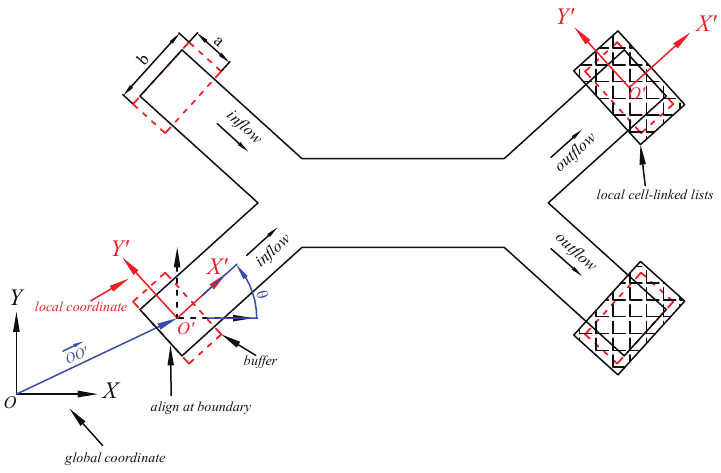}
		\caption{The rationale of arbitrary-positioned unidirectional in-/outflow buffer.}
		\label{unidirectional_buffer}
	\end{figure*}

	In Figure \ref{unidirectional_buffer}, 
	a two-dimensional complex-geometric channel	flow with multiple arbitrary-positioned in- and outlets is sketched to demonstrate the rationale
	of the arbitrary-positioned unidirectional in-/outflow buffer.
	Through the translation $ \overrightarrow{OO^{'}}= [X_{O^{'}}, Y_{O^{'}}]^{T}$ and rotation $ \theta $, the local 
	coordinate system $ O^{'}X^{'}Y^{'} $ is established at each in-/outlet, with its axis direction $\overrightarrow{O^{'}X^{'}} $ aligned to the in-/outflow direction.
	Here, $ [X_{O^{'}}, Y_{O^{'}}]^{T} $ is the position of origin $ O^{'} $ in the global coordinate system.
	The buffer, with width $ a $ and length $ b $, is centered at the origin $ O^{'} $ of the local coordinate system $ O^{'}X^{'}Y^{'} $. As shown in Figure \ref{unidirectional_buffer}, one side of the buffer is aligned with the in-/outlet boundary.
	
	In the 2-D local 
	coordinate system, the particle position $ \mathbf{r}^{'}=( X^{'}, Y^{'})^{T} $ could be obtained through the following coordinate transformation
	\begin{equation} \label{2D_transformation}
		\begin{bmatrix}
			X^{'}\\
			Y^{'}
		\end{bmatrix}
		=\begin{bmatrix}
			\cos \theta & \sin \theta\\
			-\sin \theta & \cos \theta
		\end{bmatrix}	\left(
		\begin{bmatrix}
			X\\
			Y
		\end{bmatrix}-
		\begin{bmatrix}
			X_{O^{'}}\\
			Y_{O^{'}}
		\end{bmatrix} \right),
	\end{equation}
	where $ \mathbf{r}=[X, Y]^{T} $ is the corresponding particle position in the global coordinate system, and the sign of $ \theta $ depends on the anticlockwise ($ + $) or clockwise ($ - $) rotation. 
	
	In comparison, for the coordinate transformation in the 3-D case, the unit vector $ \overrightarrow{\mathbf{n}}=\left[ n_{x}, n_{y},n_{z} \right]^{T} $ of the rotation axis is first given by normalizing the cross product of axial vectors $ \overrightarrow{OX} $ and $ \overrightarrow{O^{'}X^{'}} $ in the global coordinate system, which is a vector perpendicular to the plane spanned by $ \overrightarrow{OX} $ and $ \overrightarrow{O^{'}X^{'}} $ with the direction given by the right-hand rule
	\begin{equation} \label{rotation_axis}
		\overrightarrow{\mathbf{n}} =\dfrac{\overrightarrow{OX} \times \overrightarrow{O^{'}X^{'}}}{\left|\overrightarrow{OX} \times \overrightarrow{O^{'}X^{'}}\right|},
	\end{equation}
	and then its antisymmetric matrix $ \mathbf{S} $ can be written as 
	\begin{equation} \label{antisymmetric_matrix}
		\mathbf{S}=\begin{bmatrix}
			0 & -n_{z} & n_{y}\\
			n_{z} & 0 & -n_{x}\\
			-n_{y}&n_{x}&0
		\end{bmatrix}.
	\end{equation}

	According to the Rodrigues' rotation formula \cite{Cheng1989an, Fraiture2009his, DAI2015Euler}, the rotation matrix $ \mathbf{R} $
	is defined as
	\begin{equation} \label{rotation_matrix}
		\mathbf{R} =\mathbf{I}+\mathbf{S} \sin \theta +\mathbf{S}^{2} (1-\cos\theta), 
	\end{equation}
	where $ \mathbf{I} $ is 3$ \times $3 identity matrix, and the sign of $ \theta $ is positive.
	
	Consequently, the particle position $ \mathbf{r}^{'}=( X^{'}, Y^{'}, Z^{'})^{T} $ in the 3-D local 
	coordinate system could be given as 
	\begin{equation} \label{3D_transformation}
		\begin{bmatrix}
			X^{'}\\
			Y^{'}\\
			Z^{'}
		\end{bmatrix}
		=\mathbf{R}
		\left(
		\begin{bmatrix}
			X\\
			Y\\
			Z
		\end{bmatrix}-
		\begin{bmatrix}
			X_{O^{'}}\\
			Y_{O^{'}}\\
			Z_{O^{'}}
		\end{bmatrix} \right).
	\end{equation}

	In the 2-D inflow buffer, buffer particles will be periodically moved back to the inlet bound with $ \mathbf{r}^{'}= \mathbf{r}^{'}-\left [a,0\right ]^{T} $ \cite{Tafuni2018, Shuoguo2022free}, once they cross the buffer-fluid interface. Here, $ \mathbf{r}^{'}= \mathbf{r}^{'}-\left [a,0,0\right ]^{T} $ in the 3-D case. Consequently, due to its fixed buffer particle configuration,  buffer particles should be identified from all fluid particles in the computational domain but only for once.
	In detail, by checking $ \mathbf{r}^{'}$ of all fluid particles before the flow simulation, they will be identified as buffer particles if both $ \left|X^{'} \right|< a/2 $ and $ \left|Y^{'}\right|< b/2$ (an additional criterion of $ \left|Z^{'}\right|< c/2$ for 3-D case, with $ c $ denoting the buffer height), and then added into the set of buffer particles. Note that, each inflow buffer has its own set of buffer particles. During the subsequent flow evolution, when these identified buffer particles cross the buffer-fluid interface with 
	$ X^{'} > a/2 $ (same for 3-D case), duplicate particles are generated with the same states as the crossing ones, 
	and then treated as ordinary fluid particles in the inner flow domain.
	Meanwhile, the original buffer particles will be recycled to periodic positions at inlet bound, and their new positions in the global coordinate system are obtained through the inverse transformation of coordinates 
	\begin{equation} \label{2D_Inverse_transformation}
		\begin{bmatrix}
			X\\
			Y
		\end{bmatrix}
		=\begin{bmatrix}
			\cos \theta & \sin \theta\\
			-\sin \theta & \cos \theta
		\end{bmatrix}^{T}
		\left(
		\begin{bmatrix}
			X^{'}\\
			Y^{'}
		\end{bmatrix}-
		\begin{bmatrix}
			a\\
			0
		\end{bmatrix} \right)+
		\begin{bmatrix}
			X_{O^{'}}\\
			Y_{O^{'}}
		\end{bmatrix}.
	\end{equation}

	Correspondingly, for the 3-D case, the above inverse transformation of coordinates is modified as 
	\begin{equation} \label{3D_Inverse_transformation}
		\begin{bmatrix}
			X\\
			Y\\
			Z
		\end{bmatrix}
		=\mathbf{R}^{T}
		\left(
		\begin{bmatrix}
			X^{'}\\
			Y^{'}\\
			Z^{'}
		\end{bmatrix}-
		\begin{bmatrix}
			a\\
			0\\
			0
		\end{bmatrix} \right)+
		\begin{bmatrix}
			X_{O^{'}}\\
			Y_{O^{'}}\\
			Z_{O^{'}}
		\end{bmatrix}.
	\end{equation}

	Algorithm \ref{unidirectional_inflow_buffer} summarizes the working procedure of the unidirectional inflow buffer described above.
	\begin{algorithm}
		\label{unidirectional_inflow_buffer}
		\SetAlgoLined
		\KwData{$i$: all fluid particles in the computational domain,
			$j$: buffer particles,
			$ n $: the number of particles being checked.}
		
		\textbf{Procedure} Buffer particle identification {$ \hfill\triangleright $Execute only once}\\ 
		\For{$i=1$ \KwTo $n$}
		{
			\If{$ \left|X_{i}^{'} \right|< \frac{a}{2} $  
				\& $ \left|Y_{i}^{'}\right|< \frac{b}{2}$ for 2-D case \& $ \left|Z_{i}^{'}\right|< \frac{c}{2}$ for 3-D case}{
				add it into the set of buffer particles
			}
		}
		\textbf{Procedure} Particle generation \\
		\For{$j=1$ \KwTo $n$}
		{			
			\If{$ X_{j}^{'} > \frac{a}{2}$ for both 2-D and 3-d cases}
			{
				duplicate it, and then recycle it through the corresponding invese  transformation of coordinates Eq. \ref{2D_Inverse_transformation} or Eq. \ref{3D_Inverse_transformation}
			}
		}
		\caption{Working procedure of the arbitrary-positioned unidirectional inflow buffer.}
	\end{algorithm} 

	For the 2-D outflow buffer, as shown in Figure \ref{unidirectional_buffer},
	only the fluid particles within the nearby local cell-linked lists at each time step are checked.
	Note that, the coverage area of the local cell-linked lists is a little larger than the size of outflow buffer, approximately one layer of cells.
	If $ X^{'} >a/2$ of these checked particles (same for 3-D case), they will be directly deleted as outflow ones.
	Furthermore, due to the newly inflow fluid particles and outflow buffer particles, the buffer particle configuration of the outflow buffer necessitates an update at each time step.
	In detail, through the relabeling approach \cite{zhang2024dynamical}, at the end of each time step (after the cell-linked 
	lists have been rebuilt), if both $ \left|X^{'} \right|< a/2 $ and $ \left|Y^{'}\right|< b/2$ (an additional criterion of $ \left|Z^{'}\right|< c/2$ for 3-D case) of these checked particles, they will be relabeled as buffer particles by assigning the same integer, otherwise, they will not. Note that, each outflow buffer can have its exclusive integer for relabeling.
	Algorithm \ref{unidirectional_outflow_buffer} summarizes the working procedure of the unidirectional outflow buffer described above.

	\begin{algorithm}
		\label{unidirectional_outflow_buffer}
		\SetAlgoLined
		\KwData{$i$: fluid particles within the local cell-linked lists,
			$ n $: the number of particles being checked.}
		
		\textbf{Procedure} Particle deletion\\
		\For{$i=1$ \KwTo $n$}
		{
			\If{$ X_{i}^{'} > \frac{a}{2}$ for both 2-D and 3-d cases}
			{
				delete it			
			}	
		}
		\textbf{Procedure} Buffer particle identification (after the cell-linked 
		lists have been rebuilt at the end of present time step)\\
		\For{$i=1$ \KwTo $n$}
		{			
			\If{$ \left|X_{i}^{'} \right|< \frac{a}{2} $  
				\& $ \left|Y_{i}^{'}\right|< \frac{b}{2}$ for 2-D case \& $ \left|Z_{i}^{'}\right|< \frac{c}{2}$ for 3-D case}{
				relabel it as buffer particle
			}
		}
		\caption{Working procedure of the arbitrary-positioned unidirectional outflow buffer.}
	\end{algorithm} 

	\subsection{Bidirectional in-/outflow buffer}
	\label{bidirectional_flow_scenario}	
	\begin{figure*}  
		\includegraphics[width=\textwidth]{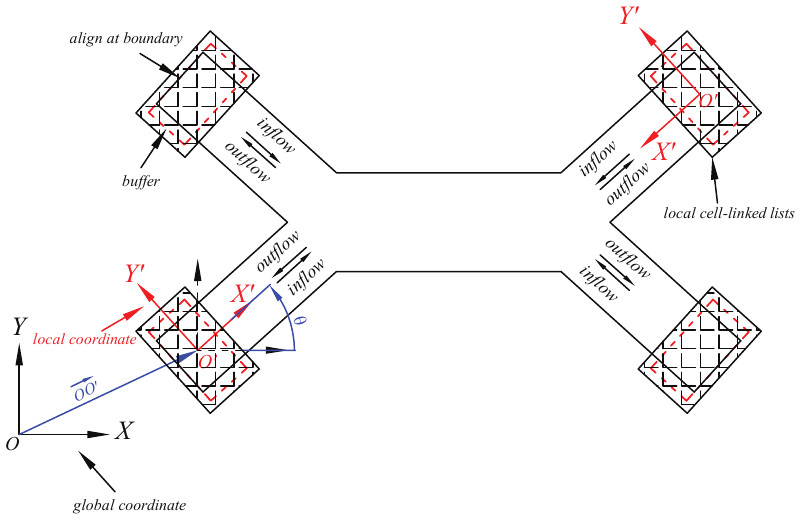} 
		\caption{The rationale of arbitrary-positioned bidirectional buffer.}
		\label{bidirectional_buffer}
	\end{figure*}

	Based on Figure \ref{unidirectional_buffer}, Figure \ref{bidirectional_buffer} comparatively illustrates the different setting and rationale of the arbitrary-positioned bidirectional buffer in the bidirectional flow.
	Here, the axis direction $\overrightarrow{O^{'}X^{'}} $ of the established local coordinate system at each in-/outlet is always aligned to the inflow direction, as shown in Figure \ref{bidirectional_buffer}.
	
	Similarly, due to the newly inflow fluid particles and outflow buffer particles, the buffer particle configuration of the bidirectional buffer should also be updated at the end of each time step in the same way of the arbitrary-positioned unidirectional outflow buffer, where the particles being checked are also restricted to those within the nearby local cell-linked lists.
	Differently, particle generation and deletion during each time step only focus on identified buffer particles. If their $ X^{'}>a/2 $ (same for 3-D case), the same procedure for particle generation as in the arbitrary-positioned unidirectional inflow buffer will be executed. Conversely, if their $ X^{'}<-a/2 $ (same for 3-D case), they will be directly deleted.
	Algorithm \ref{bidirectional_buffer_algorithm} summarizes the working procedure of the arbitrary-positioned bidirectional buffer described above.

	\begin{algorithm}
		\label{bidirectional_buffer_algorithm}
		\SetAlgoLined
		\KwData{$i$: fluid particles within the local cell-linked lists,
			$j$: buffer particles,
			$ n $: the number of particles being checked.}
		\textbf{Procedure} Particle generation and deletion \\
		\For{$j=1$ \KwTo $n$}
		{
			\If{$ X_{j}^{'} > \frac{a}{2}$ for both 2-D and 3-d cases}
			{
				duplicate it, and then recycle it through the corresponding invese  transformation of coordinates Eq. \ref{2D_Inverse_transformation} or Eq. \ref{3D_Inverse_transformation}			
			}
			
			\If{$ X_{j}^{'} <-\frac{a}{2}$ for both 2-D and 3-d cases}
			{
				delete it			
			}		
		}
		\textbf{Procedure} Buffer particle identification (after the cell-linked 
		lists have been rebuilt at the end of present time step)\\
		\For{$i=1$ \KwTo $n$}
		{			
			\If{$ \left|X_{i}^{'} \right|< \frac{a}{2} $  
				\& $ \left|Y_{i}^{'}\right|< \frac{b}{2}$ for 2-D case \& $ \left|Z_{i}^{'}\right|< \frac{c}{2}$ for 3-D case}{
				relabel it as buffer particle
			}
		}
		\caption{Working procedure of the arbitrary-positioned bidirectional buffer.}
	\end{algorithm} 
	    \section{In-/outlet boundary conditions}
	\label{In_outlet boundary conditions}	
	In this section, we introduce how to implement pressure and velocity boundary conditions at arbitrary-positioned in-/outlet through buffer particles. To cope with the truncation for density summation, the density reinitialization of Eq. (\ref{density_summation}) is not carried out for near-boundary particles (i.e., buffer particles herein).
	
	\subsection{Pressure boundary condition}
	\label{pressure_boundary_condition}	
	In the present work, the pressure boundary condition follows the work of Zhang et al. \cite{zhang2024dynamical}.
	In the global 
	coordinate system, 
	the discretized momentum equation Eq. (\ref{disctretized_momentum_conservation_equation}), when applied to buffer particles, is modified with the prescribed boundary pressure $ p_{b} $ incorporated
	\begin{equation} \label{modified_discretized_momentum_equation}
		\begin{split}
			\frac{d\mathbf{v}_{i}}{dt}=&-2\sum_{j}m_{j}(\dfrac{P^{*}_{ij}}{\rho_{i}
				\rho_{j}})\nabla W_{ij}+2p_{b}\sum_{j}(\dfrac{m_{j}}{\rho_{i}
				\rho_{j}})\nabla W_{ij}  
			\\&+2\sum_{j}m_{j}\frac{\eta\mathbf{v}_{ij} }{\rho_{i}\rho_{j}r_{ij}}
			\frac{\partial W_{ij}}{\partial r_{ij}}
		\end{split}.
	\end{equation}

	As mentioned by Holmes et al. \cite{HOLMES2021Novel}, the velocity updated by Eq. (\ref{modified_discretized_momentum_equation}) should be further projected to the normal direction $ \hat{u} $ of in-/outlet boundary in the global coordinate system 

	\begin{equation} \label{velocity_correction}
		\mathbf{v}_{i}=(\mathbf{v}_{i}\cdot \hat{u})\hat{u}. 
	\end{equation}
	Here, $ \hat{u} $ could be expressed in the 2-D case as 
	\begin{equation} \label{2D_normal_transformation}
		\hat{u}=
		\begin{bmatrix}
			\cos \theta & \sin \theta\\
			-\sin \theta & \cos \theta
		\end{bmatrix}^{T}
		\begin{bmatrix}
			1\\
			0
		\end{bmatrix}=
		\begin{bmatrix}
			\cos \theta \\
			\sin \theta
		\end{bmatrix},
	\end{equation}
	and in the 3-D case as
	\begin{equation} \label{3D_normal_transformation}
		\hat{u}=
		\mathbf{R}^{T}
		\begin{bmatrix}
			1\\
			0\\
			0
		\end{bmatrix}. 
	\end{equation}

	Furthermore, to satisfy the consistency condition at the boundary, the density of the newly populated (actually recycled) particles 
	in the buffer is obtained
	following the boundary pressure and EoS as
	\begin{equation} \label{postulate-density}
		\rho_{i} = \rho_{0} + p_{b} / c_0^2.
	\end{equation}

	\subsection{Velocity boundary condition}
	\label{Velocity_boundary_condition}	
	\begin{figure}[ht]
		\includegraphics[width=0.5\textwidth]{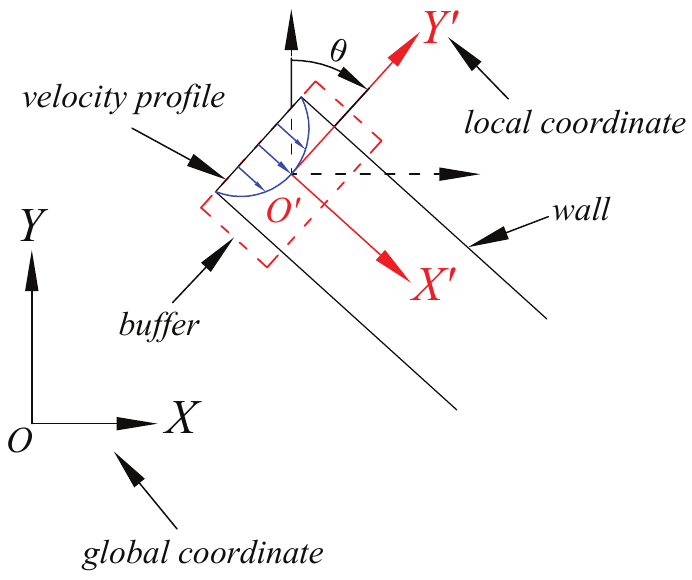} 
		\caption{The coordinate transformation on pre-defined velocity profile.}
		\label{velocity_transformation}
	\end{figure}
	For the sake of simplicity, the pre-defined velocity profile is first implemented on buffer particles within the local coordinate system, for example in Figure \ref{velocity_transformation},
	\begin{equation} \label{2D_velocity_profile}
		\begin{bmatrix}
			\mathbf{v}_{X^{'}}\\
			\mathbf{v}_{Y^{'}}
		\end{bmatrix}=
		\begin{bmatrix}
			1-(\frac{Y^{'}}{R})^{2}\\
			0
		\end{bmatrix}\qquad\text{in 2-D},
	\end{equation}
	and
	\begin{equation} \label{3D_velocity_profile}
		\begin{bmatrix}
			\mathbf{v}_{X^{'}}\\
			\mathbf{v}_{Y^{'}}\\
			\mathbf{v}_{Z^{'}}
		\end{bmatrix}=
		\begin{bmatrix}
			1-(\frac{Y^{'}}{R})^{2}\\
			0\\
			0
		\end{bmatrix}\qquad\text{in 3-D},
	\end{equation}
	where $ R $ is the channel radius.
	
	Then the corresponding velocity distribution in the global coordinate system can be derived through coordinate transformation on buffer particle velocity
	\begin{equation} \label{2D_velocity_transformation}
		\begin{bmatrix}
			\mathbf{v}_{X}\\
			\mathbf{v}_{Y}
		\end{bmatrix}
		=\begin{bmatrix}
			\cos \theta & \sin \theta\\
			-\sin \theta & \cos \theta
		\end{bmatrix}^{T}
		\begin{bmatrix}
			\mathbf{v}_{X^{'}}\\
			\mathbf{v}_{Y^{'}}
		\end{bmatrix}\qquad\text{in 2-D},
	\end{equation}
	and 
	\begin{equation} \label{3D_velocity_transformation}
		\begin{bmatrix}
			\mathbf{v}_{X}\\
			\mathbf{v}_{Y}\\
			\mathbf{v}_{Z}
		\end{bmatrix}
		=\mathbf{R}^{T}
		\begin{bmatrix}
			\mathbf{v}_{X^{'}}\\
			\mathbf{v}_{Y^{'}}\\
			\mathbf{v}_{Z^{'}}
		\end{bmatrix}\qquad\text{in 3-D}.
	\end{equation}

	The pressure boundary condition by Eq.(\ref{modified_discretized_momentum_equation}) should also be imposed at the boundary to 
	eliminate the truncated error in approximating pressure gradient, 
	but the prescribed boundary pressure $ p_{b} $ in Eq.(\ref{modified_discretized_momentum_equation}) 
	is given as $ p_{i} $, i.e., $ p_{b}=p_{i} $. Furthermore, both density
	and pressure of newly populated (actually recycled) particles in the buffer remain unchanged.
	
	\section{2-D Numerical tests}
	\label{2D_Numerical_examples}
	In this section, to validate the effectiveness of the developed arbitrary-positioned buffer in the 2-D case, several 2-D flows are simulated, including sloped-positioned VIPO and PIVO channel flows, T- and Y-shaped channel flows, and rotating sprinkler.
	For the treatment of wall boundary, a one-sided Riemann solver is employed herein \cite{article2017Chi}.
	\subsection{Sloped-positioned VIPO and PIVO channel flows}
	\label{Mixed_Poiseuille_flow}
	\begin{figure*}
		\centering     
		\includegraphics[width=0.6\textwidth]{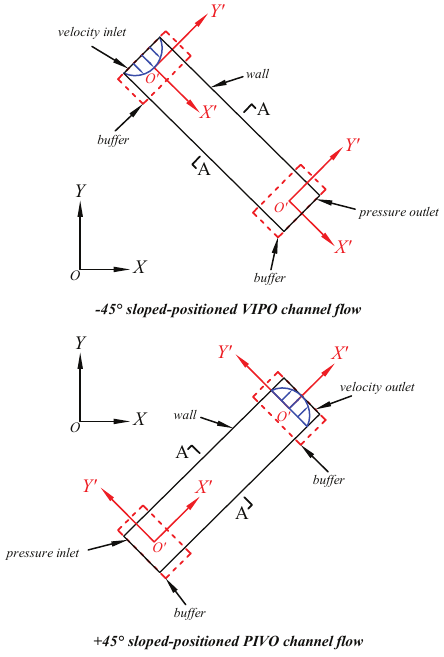}
		\caption{Schematics of sloped-positioned VIPO (top panel) and PIVO (bottom panel) channel flows. Velocity profile is extracted 
			at the cross-section $\mathbf{A-A}$ in the simulation.}
		\label{mixed_channel_flow}
	\end{figure*}
	As shown in Figure \ref{mixed_channel_flow}, 2-D sloped-positioned VIPO (Velocity Inlet, Pressure Outlet) and PIVO (Pressure Inlet, Velocity Outlet) channel flows with the available analytical solution are first simulated herein.
	At the velocity in-/outlet, the prescribed velocity profile $ \mathbf{v}_{x^{'}}(y^{'})$ in the local coordinate system is analytically
	determined by \cite{HOLMES2021Novel, Morris1997Modeling, Takeda1994Nume, Sigalotti2003SPH, Holmes2011Smoo}

	\begin{equation} \label{Poiseuille_profile}
		\mathbf{v}_{x^{'}}(y^{'}) = \frac{\mathtt{\Delta} P}{2\eta L}(\frac{d}{2}-y^{'})(\frac{d}{2}+y^{'}) ,
	\end{equation}
	where $ y^{'} \in [-\frac{d}{2},\frac{d}{2}]$ with $ d=0.001m$ denoting the distance between plates, and 
	$ \mathtt{\Delta} P=0.1 Pa $ the pressure drop across a flow length of $ L=0.004m $.
	The dynamic viscosity is set as 
	$ \eta= \sqrt{\rho d^{3} \mathtt{\Delta} P / 8L Re} $ with the fluid density 
	$ \rho=1000kg/m^{3} $ and the Reynolds number $ Re=50$. The fluid and 
	solid-wall particles are initialized on the Cartesian Lattice with a uniform particle 
	spacing of $ dp=d/50 $. 

	\begin{figure}  
		\includegraphics[width=0.45\textwidth]{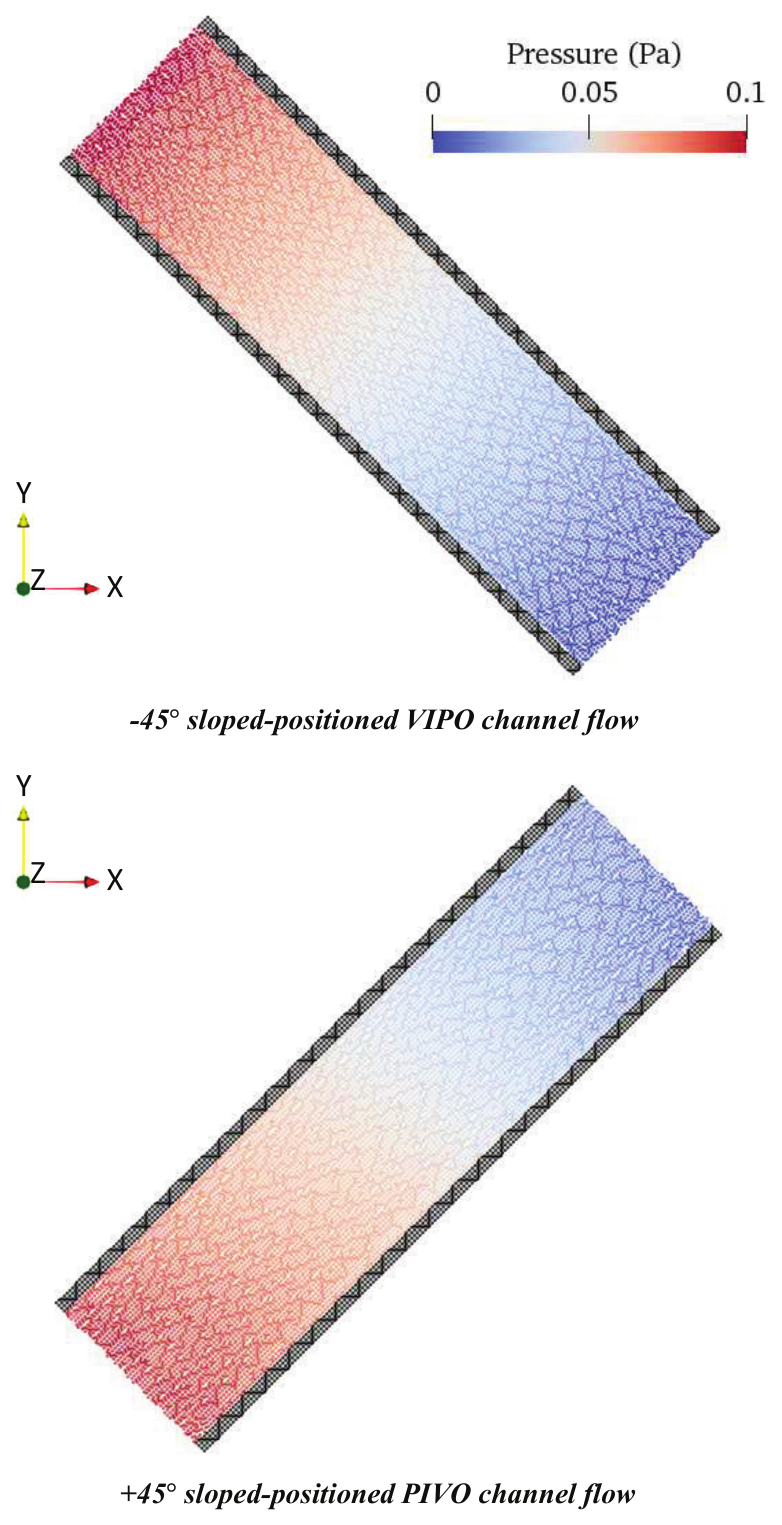}
		\caption{Pressure contours of VIPO (top panel) and PIVO (bottom panel) 
			channel flows. Time instant $ t=\infty $.}
		\label{vipo_pivo_pressure_contour}
	\end{figure}

	At the pressure outlet 
	of VIPO flow and the pressure inlet of PIVO flow,
	the prescribed boundary pressure $ p_{b} $ is set to $0 Pa$ and $0.1 Pa$, respectively.
	Consequently, in the VIPO flow, the predicted boundary pressure at the velocity inlet should 
	be $0.1 \text{ Pa}$, while at the velocity outlet of PIVO flow, 
	it should be kept at $0 \text{ Pa}$.
	Figure \ref{vipo_pivo_pressure_contour} gives the pressure contours of VIPO and VIPO flows at time $ t=\infty $. 
	It can be obviously observed that the pressure distributions align well with the above-mentioned theoretical expectations.

	\begin{figure*}
		{
			\begin{minipage}{0.49\linewidth}
				\includegraphics[width=1\textwidth]{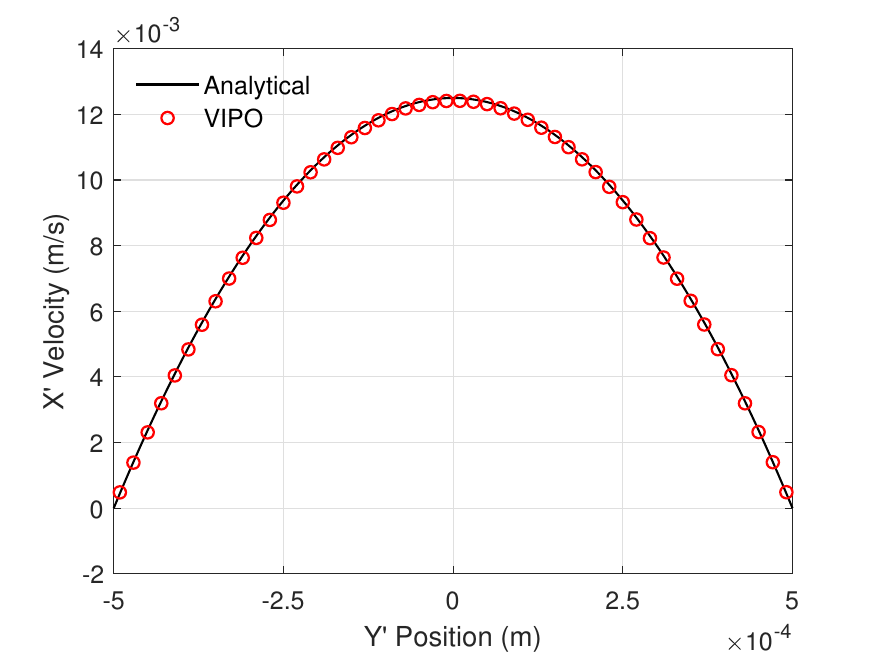}
			\end{minipage}
			\begin{minipage}{0.49\linewidth}
				\includegraphics[width=1\textwidth]{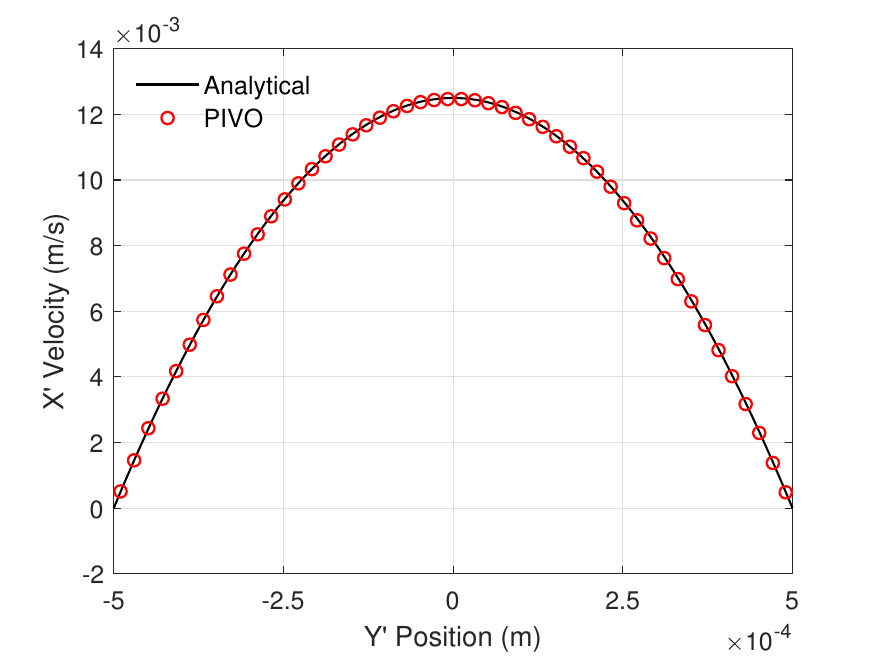}
			\end{minipage}
		}	
		\caption{Comparison of numerical and analytical velocity profiles in the local coordinate system for VIPO (left panel) and PIVO (right panel) channel flows. Time instant $ t=\infty $.
			The RMSEP errors for VIPO and PIVO channel flows are $ 1.91\% $ and $ 0.96\% $, respectively.}
		\label{PIVO_VIPO_velocity}
	\end{figure*}

	Figure \ref{PIVO_VIPO_velocity} illustrates the comparison between numerical and analytical velocity profiles for VIPO and PIVO channel flows.
	The distribution of velocity data points consistently exhibits a parabolic shape, closely aligning with the analytical solution.
	Specifically, the maximum velocity occurs at the channel's centerline, while the velocity diminishes to zero near the walls due to the no-slip boundary condition enforced on the solid walls \cite{ADAMI2012gene}.
	The numerical accuracy is further measured with the Root Mean Square Error Percentage (RMSEP) over 
	the whole velocity profile \cite{HOLMES2021Novel}
	\begin{equation} \label{RMSEP}
		RMSEP =\sqrt{\dfrac{1}{N}\sum_{n=1}^N \left(\dfrac{\mathbf{v}_{x^{'}}(y^{'}_{n})
				-\mathbf{\widetilde{v}}_{x^{'}}(y^{'}_{n})}
			{\mathbf{v}_{x^{'}}(y^{'}_{n})}\right)^{2}},
	\end{equation}
	where $ N $ is the number of measuring points, 
	$ \mathbf{v}_{x^{'}}(y^{'}_{n}) $ and $\mathbf{\widetilde{v}}_{x^{'}}(y^{'}_{n}) $ 
	the analytical and numerical velocities, respectively. 
	The RMSEP errors of VIPO and PIVO channel flows are $1.91 \%$ and $0.96 \%$, 
	respectively, demonstrating a high level of concordance with the analytical solution.

	\subsection{T- and Y-shaped channel flows}
	\label{TandY_flow} 
\begin{figure*}     
		\includegraphics[width=0.9\textwidth]{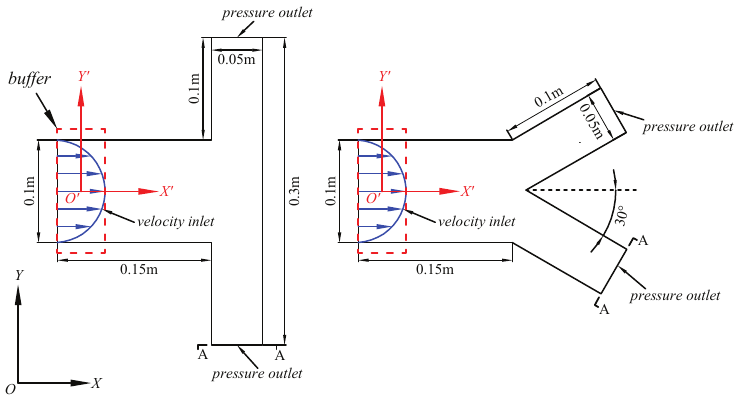}
		\caption{T- and Y-shaped channel flow schematics with the highlighted unidirectional inflow buffer. Velocity profile is extracted 
			at the cross-section $\mathbf{A-A}$ in the simulation.}		
		\label{TYschematic}
	\end{figure*}

	Here we conduct two preliminary 2-D application cases where daughter pipes are connected to the main pipe at different angles, i.e., T- and Y-shaped channel flows in Figure \ref{TYschematic}, and compare them to the results obtained from Fluent simulations.
	In the local coordinate system, the inflow velocity profile is given as $ \mathbf{v}_{x^{'}} =400(0.05^{2}-(y^{'})^{2})$ with the 
	characteristic velocity $ U_{max}=1m/s $. Hence, the dynamic viscosity $ \eta= 0.1 \rho U_{max}/Re $ with the Reynolds number $ Re=100 $ and fluid density $ \rho=1000kg/m^{3} $. The boundary pressure $ p_{b} $ at the pressure outlet is set to $ 0pa$. The uniform particle/grid resolution $ dp $ is $ 10^{-3}m $.

	\begin{figure}
		\includegraphics[width=0.7\textwidth]{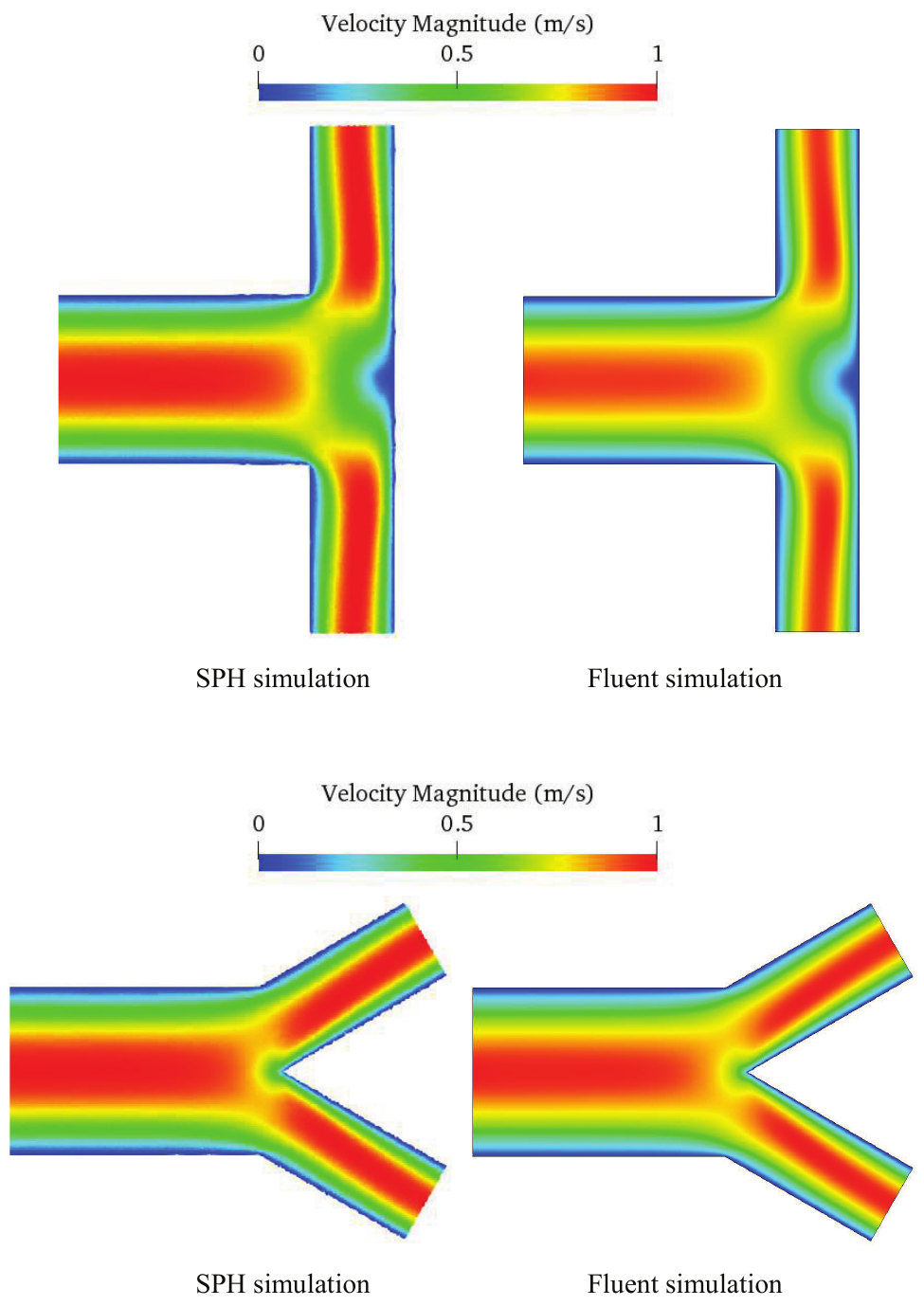}
		\caption{Qualitative comparison of velocity contours from SPH and Fluent simulations for T- (top panel) and Y- (bottom panel) shaped channel flows. Time instant $ t=\infty $.}
		\label{TY_velocity_contour}
	\end{figure}

\begin{figure*}
		{
			\begin{minipage}{0.49\linewidth}
				\includegraphics[width=1\textwidth]{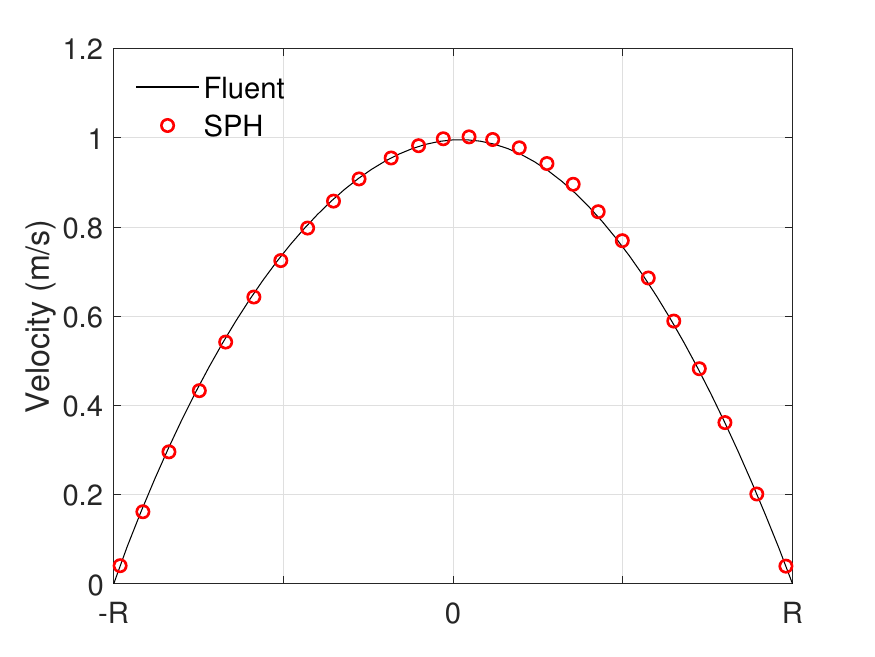}
			\end{minipage}
			\begin{minipage}{0.49\linewidth}
				\includegraphics[width=1\textwidth]{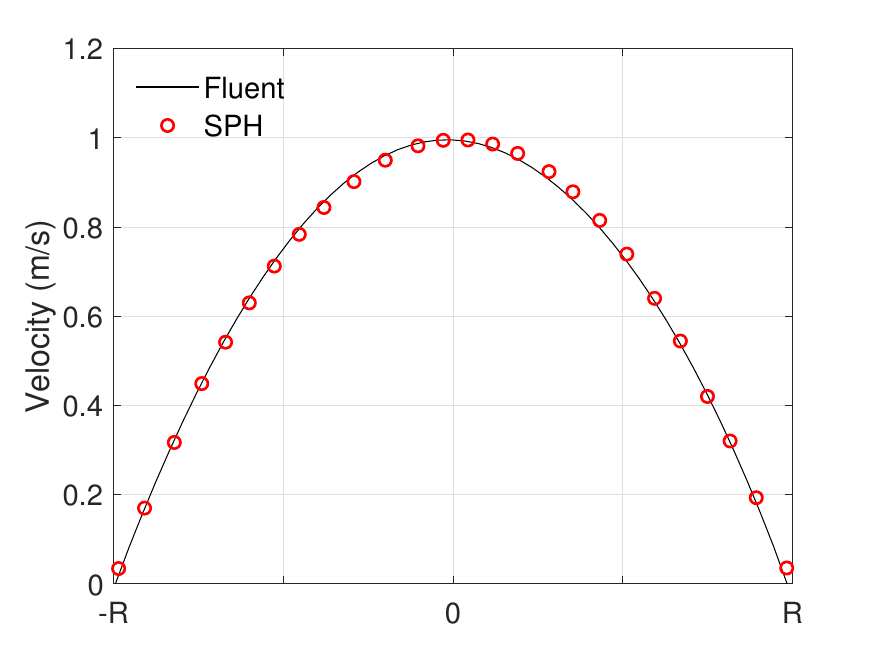}
			\end{minipage}
		}	
		\caption{Comparison of SPH and Fluent velocity profiles in the local coordinate system for T- 
			(left panel) and Y- (right panel) shaped channel flows. Time instant $ t=\infty $. $ R $ denotes the radius of daughter channel.
			The RMSEP errors for T- and Y-shaped channel flows are $ 2.29\% $ and $ 3\% $, respectively.}
		\label{TY_velocity_compare}
	\end{figure*}

	The left panels of Figure \ref{TY_velocity_contour} illustrate the SPH velocity contours of T- and Y-shaped channel flows at time $ t =\infty$, respectively.  
	At the fork junction, i.e., the stagnation point, the fluid velocity decreases to almost zero, especially in the T-shaped channel flow.
	The right panels of Figure \ref{TY_velocity_contour} show the corresponding results of Fluent simulation with the same model settings. 
	Although good agreement is already demonstrated in the qualitative comparison of Figure \ref{TY_velocity_contour}, we further conduct a quantitative comparison with velocity profiles extracted at the outlet, as shown in Figure \ref{TY_velocity_compare}.
	Both the shapes of velocity profiles from SPH and Fluent simulations present consistent parabolic forms, with nearly identical velocity magnitudes. 
	By replacing the analytical velocity $ \mathbf{v}_{x^{'}}(y^{'}_{n}) $ in Eq. (\ref{RMSEP}) as the numerical velocity from the Fluent simulation, the RMSEP errors of T- and Y-shaped channel flows are measured as $ 2.29\% $ and $ 3.01\% $, respectively, 
	which are considered acceptable.

	\subsection{Rotating sprinkler}
\begin{figure}    
		\includegraphics[width=0.75\textwidth]{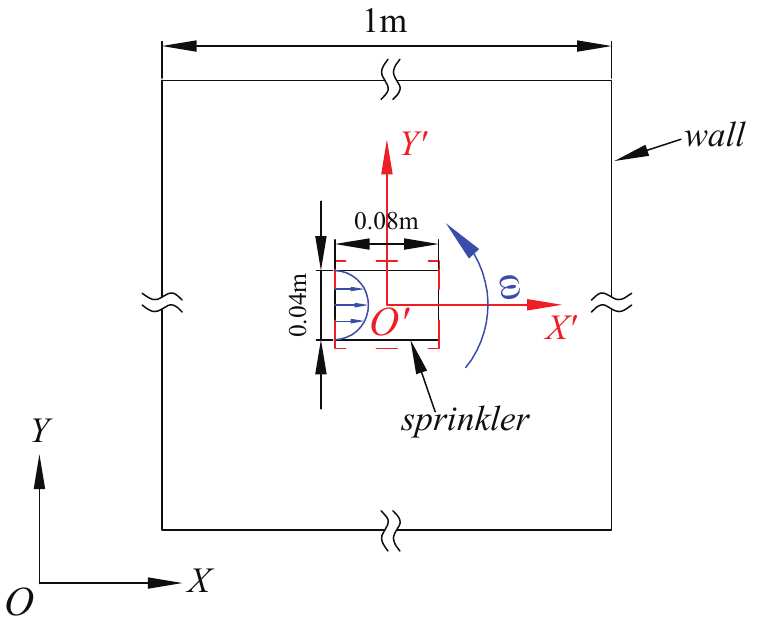}
		\caption{Top view of rotating sprinkler. The sprinkler consisting of buffer particles is rotating with the constant angular velocity $ \omega $.}
		\label{rotating_sprinkler_schematic}
	\end{figure}

	To better demonstrate the versatility of the developed arbitrary-positioned buffer in engineering applications, the 2-D rotating sprinkler, referring to a lawn sprinkler, is simulated here, which is realized by rotating buffer particles with a constant angular velocity $ \omega $, as depicted in Figure \ref{rotating_sprinkler_schematic}.
	In this top view simulation, gravity $ \mathbf{g} $ is ignored, and the fluid is considered inviscid with the density of $ \rho=1000kg/m^{3} $.
	The inflow velocity profile in the sprinkler is defined within the local coordinate system as $ \mathbf{v}_{x^{'}} =2500(0.02^{2}-(y^{'})^{2})$.
	The uniform particle spacing is $ dp=0.005m $.

	\begin{figure}
		\includegraphics[width=0.8\textwidth]{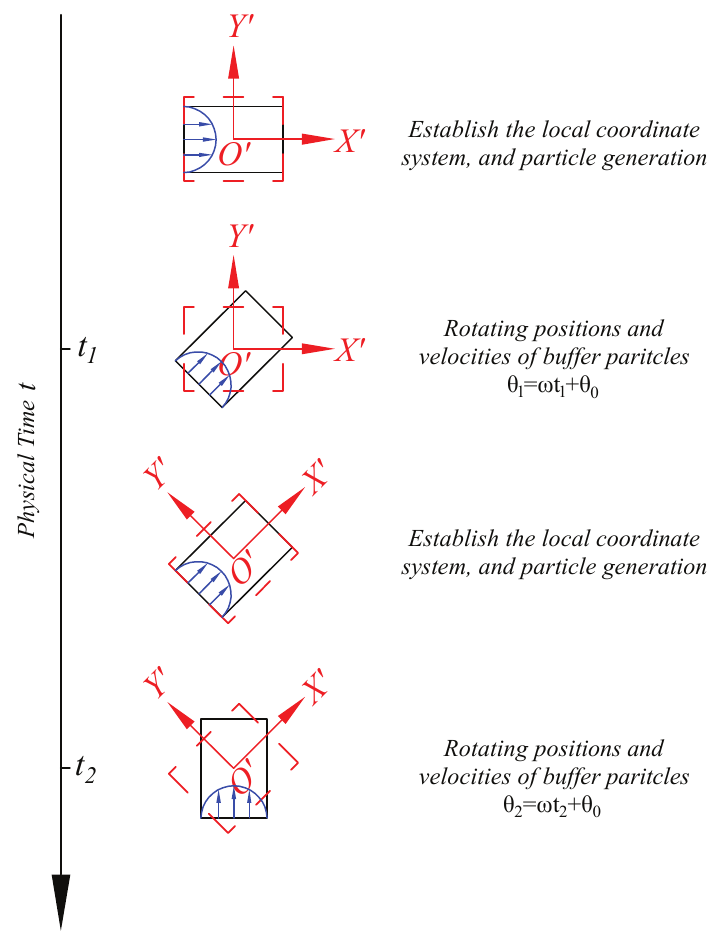}
		\caption{The flow chart on rotating the sprinkler through coordinate transformation.}
		\label{rotation_procedure}
	\end{figure}

	Figure \ref{rotation_procedure} illustrates the flowchart for rotating the sprinkler through coordinate transformation. The rotation angle $ \theta $ is given as $ \theta=\omega t+\theta_{0} $, where $ \theta_{0} $ denotes the initial rotation angle at time $ t=0 $. In the present case, the angular velocity $ \omega =2 rad/s$ and $ \theta_{0}=0 $. 
	Before rotating the sprinkler at $ t_{1} $, 
	the particle generation in this unidirectional inflow buffer has already been completed in the established local coordinate system. Then, based on this local coordinate system, the positions and velocities of buffer particles are transformed to new ones through coordinate rotation at $ t_{1} $. After this rotation, the local coordinate system is reestablished according to the new position of sprinkler, preparing it for particle generation in next time step. Similarly, based on this newly established local coordinate system, the sprinkler undergoes the next rotation at $ t_{2} $, following the same procedure as at $ t_{1} $. Thus, the iterative cycle continues. 

	\begin{figure*} 
		\includegraphics[width=0.93\textwidth]{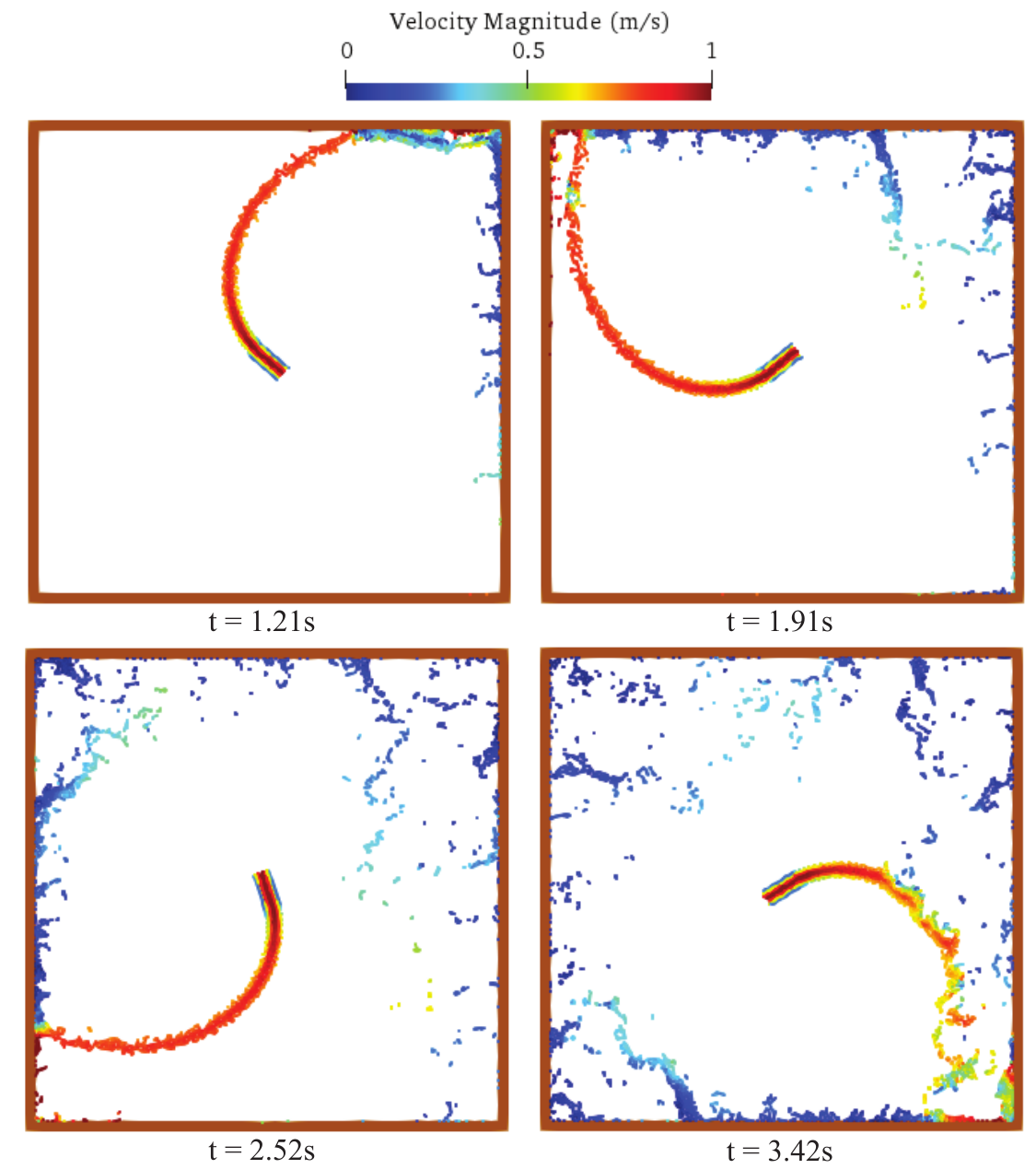}
		\caption{Velocity contours of rotating sprinkler at different time instants.}
		\label{sprinkler_velocity_contour}
	\end{figure*}

	Figure \ref{sprinkler_velocity_contour} shows the velocity contours at different time instants. During the continuous rotation, the velocity profile within the sprinkler consistently maintains the prescribed parabolic shape, and new inflow particles at the sprinkler’s outlet are effectively generated. Due to inertia, the fluid bulk exhibits a curved shape, consistent with physical principles. At these four instants, all rotated angles of the sprinkler match the predicted angles based on the angular velocity. All these flow behaviors demonstrate the good reliability and performance of the arbitrary-positioned buffer developed by introducing the coordinate transformation.
	
	\section{3-D Numerical tests}
	\label{3D_Numerical_examples}
	In this section, to validate the effectiveness of the developed arbitrary-positioned buffer in the 3-D case, two 3-D flows with the same wall boundary condition as the 2-D flow are simulated, including the sloped-positioned pulsatile pipe flow and aorta flow.
	\subsection{Sloped-positioned pulsatile pipe flow}
	\label{Pulsatile_flow}
\begin{figure*}
		\begin{adjustbox}
			{addcode={\begin{minipage}{\width}}{\caption{The pressure gradient as a cosine function 
							(upper panel) in the $ -45^{\circ} $ sloped-positioned pulsatile pipe flow, and the comparison (lower panel) of 
							numerical and analytical velocity profiles in the local coordinate system at 13 indicative instants. 
							The horizontal grid dimension in the lower panel spans a velocity range 
							from $ 0 $ to 5.0E-3 $m/s $. The errors at 13 instants, from 
							$ t=2\pi $ to $ 4\pi $, are $ 1.43\% $, $ 2.27\% $, 
							$ 2.8\% $, $ 3.45\% $, $ 3.29\% $, $ 2.38\% $, $ 2.91\% $, 
							$ 2.65\% $, $ 2.89\% $, $ 3.55\% $, $ 2.43\% $, $ 1.85\% $ and $ 1.35\% $,  respectively.}\label{fig:Pulsatile_flow_velocity_comparison}\end{minipage}},
				rotate=90,center}   		
			\includegraphics[width=0.6\textwidth,angle=-90]{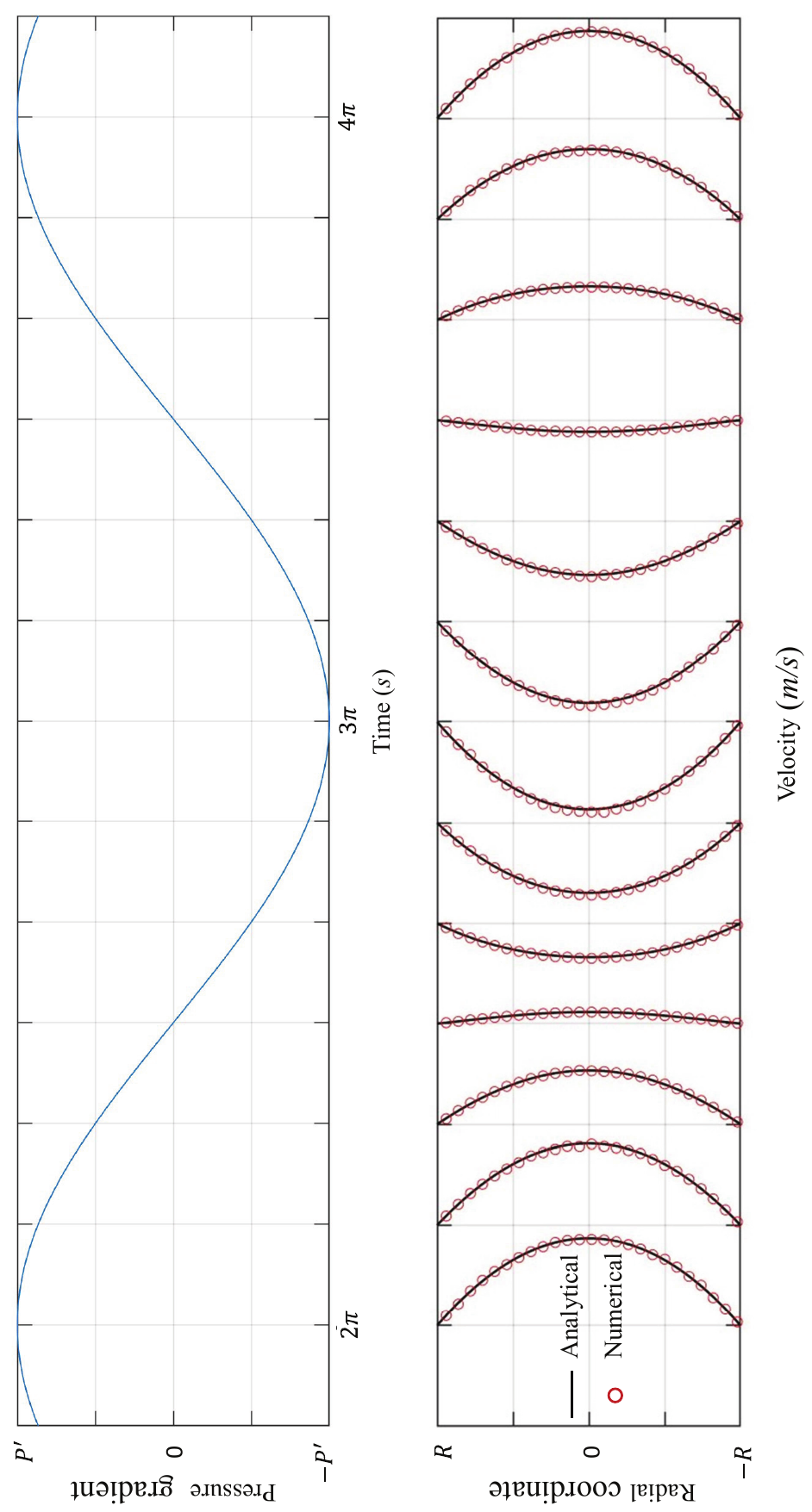}   		
		\end{adjustbox} 
	\end{figure*}

	Building upon the 2-D sloped-positioned VIPO channel flow in section \ref{Mixed_Poiseuille_flow}, an additional dimension, typically representing the radial direction in the 3-D case, has been included to model the 3-D sloped-positioned pulsatile pipe flow, while all other physical and geometrical parameters remain unchanged.
	Differently, the pressure boundary condition is imposed at both in- and outlets. When the constant pressure gradient is modified 
	to follow a cosine function over time, i.e., $P^{'}=\mathtt{\Delta} P/L=0.1\cos(t)/0.004 
	=25\cos(t)$, as illustrated in the upper panel of 
	Figure \ref{fig:Pulsatile_flow_velocity_comparison}, a bidirectional pulsatile flow is facilitated.
	The analytical velocity profile in the local coordinate system, i.e., $\mathbf{v}_{x^{'}}(r,t)$ at time $t$ and radius $r$ from the pipe axis, could be given as \cite{WOMERSLEY1955met}

	\begin{equation} \label{Pulsatile_flow_profile}
		\mathbf{v}_{x^{'}}(r,t) = \mathbf{Re}\bigg\{\sum_{n=0}^N \dfrac{i P_{n}^{'}}
		{\rho n \omega}\bigg[1-\dfrac{J_{0}(\alpha n^{\frac{1}{2}}i^{\frac{3}{2}}
			\frac{r}{R})}{J_{0}(\alpha n^{\frac{1}{2}}i^{\frac{3}{2}})}\bigg]e^
		{in\omega t}\bigg\}, 
	\end{equation}
	where $ r \in (0,R)$, $ \alpha=R\sqrt{\omega \rho/\eta} $ the dimensionless 
	Womersley number, $ \omega $ the angular frequency of the first harmonic of 
	a Fourier series of an oscillatory pressure gradient, $ P_{n}^{'} $ the pressure 
	gradient magnitude for the frequency $ n\omega $, $ J_{0}(\cdot) $ the Bessel 
	function of first kind and order zero, $ i $ the imaginary number, and 
	$ \mathbf{Re}\left\{\cdot\right\}$ the real part of a complex number.
	
	In the present pulsatile pipe flow, the Womersley number $ \alpha = 0.8409$ is less than 2, which 
	determines a parabolic shape of the pulsatile flow profile. 
	Moreover, during the time evolution, consistent with the cosine-varying 
	pressure gradient, both velocity magnitude and 
	direction exhibit periodicity.
	The lower panel of Figure \ref{fig:Pulsatile_flow_velocity_comparison} correspondingly shows this periodic evolution of the parabolic shape by extracting velocity profiles at 13 instants between $ t = 2\pi $ and $ 4\pi $. During the initial quarter, the parabolic velocity profile maintains a forward flow direction, but with a diminishing magnitude over time. The flow undergoes a reversal at $ t = 2.5\pi $, followed by gradual acceleration until reaching the maximum velocity at $ t = 3\pi $. In the latter half of the period, this flow behavior repeats, but with the opposite evolution of flow direction. Moreover, employing RMSEP for accuracy assessment reveals errors of approximately $3\%$ across all 13 time instants, demonstrating a robust agreement with their corresponding analytical solutions.
	
\begin{figure}   
		\includegraphics[width=0.45\textwidth]{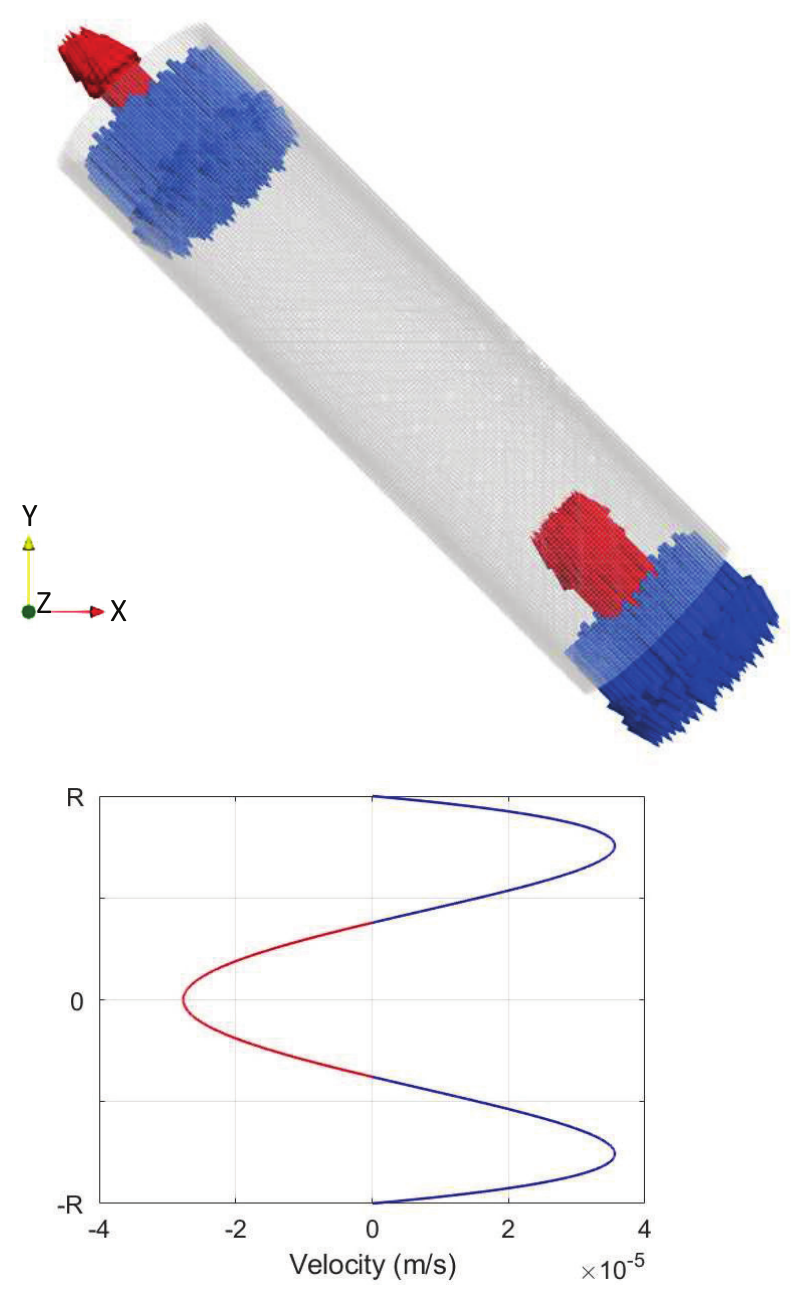}
		\caption{Velocity profile with mixed forward and reverse flows at the in-/outlet 
			boundary (top panel) of $ -45^{\circ} $ sloped-positioned pulsatile pipe flow, corresponding to the analytical velocity profile 
			(bottom panel) in the local coordinate system
			at time instant $ t=11.116s $. 
			The velocity 
			vector in top panel is scaled to a uniform length, 
			with the flow direction indicated by the arrow.}
		\label{bidirectional_generation_deletion}
	\end{figure}

	Furthermore, aligning with the analytical prediction of mixed in- and outflows at time $ t=11.116s $ shown in the bottom panel of Figure \ref{bidirectional_generation_deletion}, the top panel of Figure \ref{bidirectional_generation_deletion} depicts the mixed forward and reverse flows occurring at the in-/outlet boundary of the practical pulsatile pipe flow. This demonstrates the good performance of the arbitrary-positioned bidirectional buffer in managing particle generation and deletion.


	\subsection{Aorta flow}
	\label{Aorta_flow}
\begin{figure}
		\includegraphics[width=0.5\textwidth]{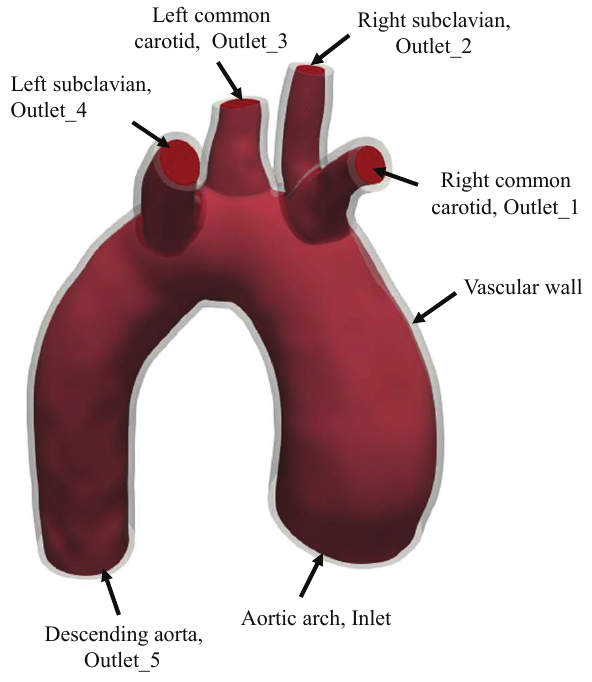}
		\caption{Schematic of the aorta flow.}
		\label{vessel_schematic}
	\end{figure}

	To investigate the feasibility of the developed arbitrary-positioned buffer for handling practical 3-D complex flows with multiple arbitrary-positioned in- and outlets, a preliminary flow simulation is conducted on an aorta model, 
	as shown in Figure \ref{vessel_schematic}. 
	The vascular wall is treated as a rigid body, while the blood is characterized as a Newtonian fluid with a density of 1060 $ kg/m^{3} $ \cite{Katia2005Multi} and a viscosity of 0.00355 $ Pa\cdot s $ \cite{Gallo2012Vivo}.

\begin{figure*} 
		\begin{adjustbox}
			{addcode={\begin{minipage}{\width}}{\caption{The time history 
							of the inflow velocity and all outlet relative pressure.}
						\label{fig:Aorta_relative_pressure}\end{minipage}},rotate=90,center}   		
			\includegraphics[width=1.4\textwidth,angle=0]{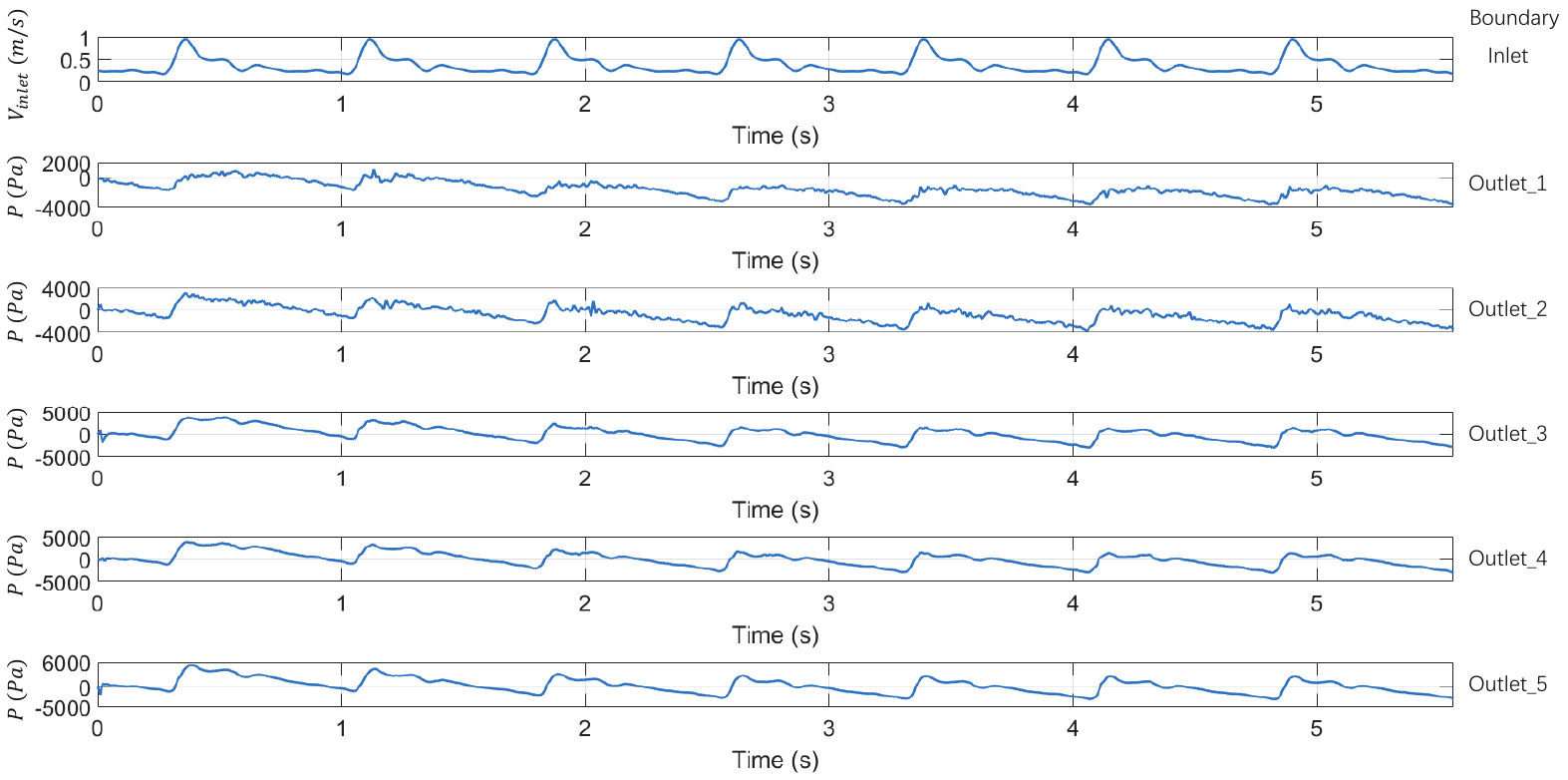}   		
		\end{adjustbox} 
	\end{figure*}

	At the inlet, a velocity boundary condition featuring a plug velocity profile is enforced to facilitate unidirectional inflow. However, the velocity magnitude exhibits pulsatility, as depicted in the top panel of Figure \ref{fig:Aorta_relative_pressure}, which can be approximated as Fourier series in term of time \cite{Taib2019blood}
	\begin{equation} \label{Fourier_Series}
		\mathbf{v}_{inlet}(t)= a_{0}+\sum_{n=1}^8 a_{n}cos(nt\omega)+
		\sum_{n=1}^8 b_{n}sin(nt\omega),
	\end{equation} 
	where the empirical variables for the Fourier series are listed in 
	Table \ref{empirical_variables}.   
	\begin{table}
		\caption{The empirical variables for the Fourier series \cite{Taib2019blood}}
        \begin{ruledtabular}
			\begin{tabular}{cccc}
				Parameter  & \multicolumn{1}{c}{Value} & \multicolumn{1}{c}{Parameter} 
				& \multicolumn{1}{c}{Value} \\
				\hline
				$ a_{0} $ & 0.3782 & $ \omega $ & 8.302 \\
				$ a_{1} $ & -0.1812 & $ b_{1} $ & -0.07725 \\
				$ a_{2} $ & 0.1276 & $ b_{2} $ & 0.01466 \\
				$ a_{3} $ & -0.08981 & $ b_{3} $ & 0.04295 \\
				$ a_{4} $ & 0.04347 & $ b_{4} $ & -0.06679 \\
				$ a_{5} $ & -0.05412 & $ b_{5} $ & 0.05679 \\
				$ a_{6} $ & 0.02642 & $ b_{6} $ & -0.01878 \\
				$ a_{7} $ & 0.008946 & $ b_{7} $ & 0.01869 \\
				$ a_{8} $ & -0.009005 & $ b_{7} $ & -0.01888 \\			    
		\end{tabular}
	    \end{ruledtabular}
		\label{empirical_variables} 
	\end{table}

\begin{figure*} 
		\includegraphics[width=0.9\textwidth]{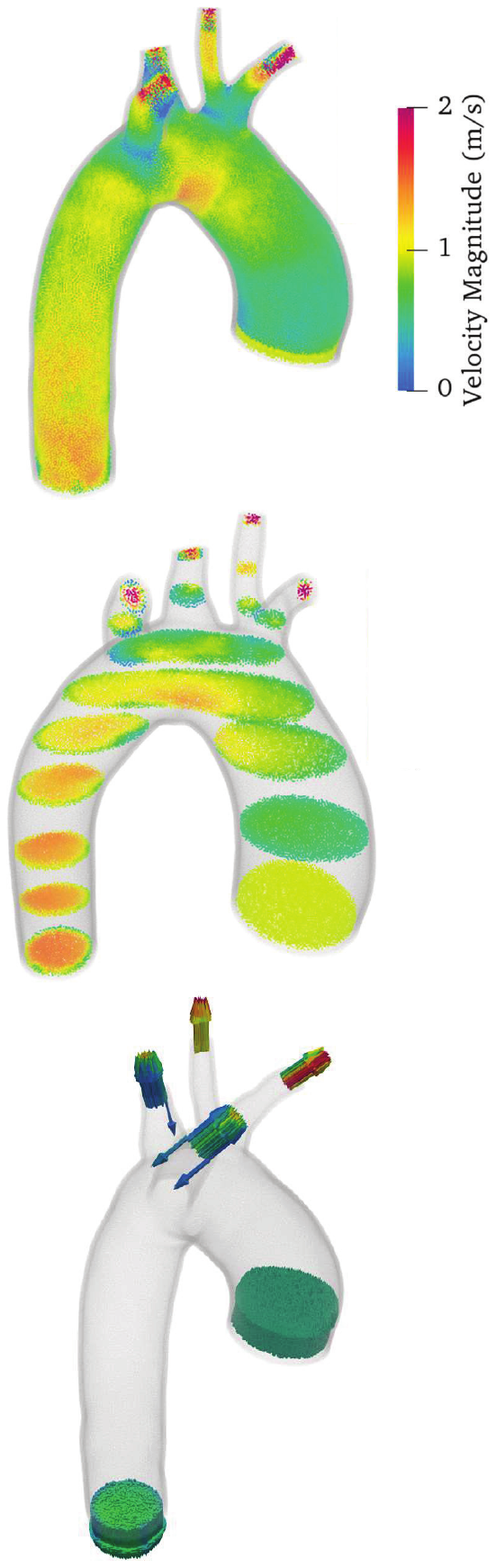}
		\caption{Velocity contour and slices of aorta flow during the peak systole, along with indicating corresponding velocity vectors of boundary particles. Time instant $ t=4.9s $. The velocity 
			vector in bottom panel is scaled to a uniform length, 
			with the flow direction indicated by the arrow.}
		\label{Aorta_velocity_contour}
	\end{figure*}

	At all outlets, the pressure boundary condition is applied by employing the time-dependent boundary pressure $ P(t) $, thus the bidirectional flow happens at outlets during the cyclical 
	blood flow, as shown in bottom panel of Figure \ref{Aorta_velocity_contour}.  
	Here, the 3-element (RCR) Windkessel model \cite{Catanho2012Model} is employed for the real-time computing of $ P(t) $

	\begin{equation} \label{windkessel}
		(1+\dfrac{R_{p}}{R_{d}})Q(t)+CR_{p}\frac{dQ(t)}{dt}=\dfrac{P(t)}
		{R_{d}}+C\dfrac{dP(t)}{dt},
	\end{equation} 
	where $ Q(t) $ is the blood flow volume,  $ C $ the arterial compliance, 
	$ R_{p} $ the characteristic resistance, and $ R_{d} $ the peripheral resistance. 
	Table \ref{Windkessel_parameters} lists the values of Windkessel parameters \cite{Kim2007coupling, Sudharsan2018effect} for allocation across diverse daughter vessels.

	\begin{table*}
		\caption{Parameters for the Windkessel outlet boundary 
			conditions \cite{Kim2007coupling, Sudharsan2018effect}}
	\begin{ruledtabular}
			\begin{tabular}{cccc}
				Vessel  & \multicolumn{1}{c}{\textbf{$ R_{p}(dynes\: s/cm^{5}) $}} 
				& \multicolumn{1}{c}{\textbf{$ C(cm^{5}/dynes) $}} & \multicolumn{1}{c}
				{\textbf{$ R_{d}(dynes\: s/cm^{5})$}} \\
				\hline 
				Right common carotid &  1180     &  7.70E-5  &  18400      \\
				Right subclavian&     1040      &  8.74E-5        &     16300     \\
				Left common carotid&   1180     & 7.70E-5 &      18400 \\
				Left subclavian&      970     & 9.34E-5   &     15200     \\
				Descending aorta&    188      &    4.82E-4      &     2950      \\  			    
		\end{tabular}
	\end{ruledtabular}
		\label{Windkessel_parameters} 
	\end{table*}

	The top and middle panels of Figure \ref{Aorta_velocity_contour} shows the velocity contour and slices during peak systole. In the present plug-inlet boundary condition, the velocity peak isn't always at the center of the cross-section. Especially in the descending aorta, due to inertia, the flow inside the curved vessel is pushed towards the outer side of the arch, which is in good agreement with the observations in literature \cite{Sudharsan2018effect}. Furthermore, as depicted in the bottom panel of Figure \ref{Aorta_velocity_contour}, which indicates the corresponding velocity vectors of boundary particles, the present arbitrary-positioned bidirectional buffer can effectively handle particle generation and deletion at these typical mixed in- and outflow boundaries. 

	Figure \ref{fig:Aorta_relative_pressure} illustrates the temporal evolution of relative pressure at all outlets across 7 cardiac cycles. Notably, all observed outlet relative pressure waveforms exhibit a characteristic folding line shape, consistent with established analytical and numerical findings in literature \cite{Catanho2012Model}. It is pertinent to acknowledge that, relative to real physiological conditions, the parameters governing inflow velocity, Windkessel parameters of all daughter vessels, and the geometric specifications of the aorta stem from disparate data sources, thereby lacking strict alignment. Moreover, the physiological pressure data corresponding to the present aorta model are also unavailable. Consequently, our primary emphasis lies in assessing the fidelity of the blood pressure waveform rather than pressure magnitudes.
	
	\section{Conclusion}
	\label{Conclusion}
	In this paper, we develop a generalized and high-efficiency arbitrary-positioned uni-/bidirectional
	buffer for SPH method.
	By employing coordinate transformation, the position comparison for particle generation and deletion is conducted in the local coordinate system established at each in-/outlet. Since one local axis is located perpendicular to the in-/outlet boundary, the position comparison can be simplified and conducted only in this coordinate dimension.
	To further enhance computational efficiency, the particle candidates subjected to position comparison are restricted to those within the local cell-linked lists nearby the defined buffer zone.
	The numerical results of 2-D and 3-D flows demonstrate the excellent effectiveness and versatility of the developed buffer for particle generation and deletion at arbitrary-positioned in- and outlets. As the future work, we will implement the developed buffer in various free-surface open-channel flows with multiple arbitrary-positioned in- and outlets.

	\nocite{*}
	\bibliography{reference_literature}

\begin{thebibliography}{52}%
\makeatletter
\providecommand \@ifxundefined [1]{%
 \@ifx{#1\undefined}
}%
\providecommand \@ifnum [1]{%
 \ifnum #1\expandafter \@firstoftwo
 \else \expandafter \@secondoftwo
 \fi
}%
\providecommand \@ifx [1]{%
 \ifx #1\expandafter \@firstoftwo
 \else \expandafter \@secondoftwo
 \fi
}%
\providecommand \natexlab [1]{#1}%
\providecommand \enquote  [1]{``#1''}%
\providecommand \bibnamefont  [1]{#1}%
\providecommand \bibfnamefont [1]{#1}%
\providecommand \citenamefont [1]{#1}%
\providecommand \href@noop [0]{\@secondoftwo}%
\providecommand \href [0]{\begingroup \@sanitize@url \@href}%
\providecommand \@href[1]{\@@startlink{#1}\@@href}%
\providecommand \@@href[1]{\endgroup#1\@@endlink}%
\providecommand \@sanitize@url [0]{\catcode `\\12\catcode `\$12\catcode
  `\&12\catcode `\#12\catcode `\^12\catcode `\_12\catcode `\%12\relax}%
\providecommand \@@startlink[1]{}%
\providecommand \@@endlink[0]{}%
\providecommand \url  [0]{\begingroup\@sanitize@url \@url }%
\providecommand \@url [1]{\endgroup\@href {#1}{\urlprefix }}%
\providecommand \urlprefix  [0]{URL }%
\providecommand \Eprint [0]{\href }%
\providecommand \doibase [0]{http://dx.doi.org/}%
\providecommand \selectlanguage [0]{\@gobble}%
\providecommand \bibinfo  [0]{\@secondoftwo}%
\providecommand \bibfield  [0]{\@secondoftwo}%
\providecommand \translation [1]{[#1]}%
\providecommand \BibitemOpen [0]{}%
\providecommand \bibitemStop [0]{}%
\providecommand \bibitemNoStop [0]{.\EOS\space}%
\providecommand \EOS [0]{\spacefactor3000\relax}%
\providecommand \BibitemShut  [1]{\csname bibitem#1\endcsname}%
\let\auto@bib@innerbib\@empty
\bibitem [{\citenamefont {Holmes}\ and\ \citenamefont
  {Pivonka}(2021)}]{HOLMES2021Novel}%
  \BibitemOpen
  \bibfield  {author} {\bibinfo {author} {\bibfnamefont {D.~W.}\ \bibnamefont
  {Holmes}}\ and\ \bibinfo {author} {\bibfnamefont {P.}~\bibnamefont
  {Pivonka}},\ }\bibfield  {title} {\enquote {\bibinfo {title} {Novel pressure
  inlet and outlet boundary conditions for smoothed particle hydrodynamics,
  applied to real problems in porous media flow},}\ }\href@noop {} {\bibfield
  {journal} {\bibinfo  {journal} {Journal of Computational Physics}\ }\textbf
  {\bibinfo {volume} {429}},\ \bibinfo {pages} {110029} (\bibinfo {year}
  {2021})}\BibitemShut {NoStop}%
\bibitem [{\citenamefont {Braun}\ \emph {et~al.}(2015)\citenamefont {Braun},
  \citenamefont {Wieth}, \citenamefont {Koch},\ and\ \citenamefont
  {Bauer}}]{Braun2015}%
  \BibitemOpen
  \bibfield  {author} {\bibinfo {author} {\bibfnamefont {S.}~\bibnamefont
  {Braun}}, \bibinfo {author} {\bibfnamefont {L.}~\bibnamefont {Wieth}},
  \bibinfo {author} {\bibfnamefont {R.}~\bibnamefont {Koch}}, \ and\ \bibinfo
  {author} {\bibfnamefont {H.}~\bibnamefont {Bauer}},\ }\bibfield  {title}
  {\enquote {\bibinfo {title} {{A framework for permeable boundary conditions
  in SPH: Inlet, Outlet, Periodicity}},}\ \ }(\bibinfo {year}
  {2015})\BibitemShut {NoStop}%
\bibitem [{\citenamefont {Negi}\ and\ \citenamefont
  {Ramachandran}(2022{\natexlab{a}})}]{Negi2022how}%
  \BibitemOpen
  \bibfield  {author} {\bibinfo {author} {\bibfnamefont {P.}~\bibnamefont
  {Negi}}\ and\ \bibinfo {author} {\bibfnamefont {P.}~\bibnamefont
  {Ramachandran}},\ }\bibfield  {title} {\enquote {\bibinfo {title} {How to
  train your solver: Verification of boundary conditions for smoothed particle
  hydrodynamics},}\ }\href@noop {} {\bibfield  {journal} {\bibinfo  {journal}
  {Physics of Fluids}\ }\textbf {\bibinfo {volume} {34}} (\bibinfo {year}
  {2022}{\natexlab{a}})}\BibitemShut {NoStop}%
\bibitem [{\citenamefont {Federico}\ \emph {et~al.}(2012)\citenamefont
  {Federico}, \citenamefont {Marrone}, \citenamefont {Colagrossi},
  \citenamefont {Aristodemo},\ and\ \citenamefont
  {Antuono}}]{FEDERICO2012Simula}%
  \BibitemOpen
  \bibfield  {author} {\bibinfo {author} {\bibfnamefont {I.}~\bibnamefont
  {Federico}}, \bibinfo {author} {\bibfnamefont {S.}~\bibnamefont {Marrone}},
  \bibinfo {author} {\bibfnamefont {A.}~\bibnamefont {Colagrossi}}, \bibinfo
  {author} {\bibfnamefont {F.}~\bibnamefont {Aristodemo}}, \ and\ \bibinfo
  {author} {\bibfnamefont {M.}~\bibnamefont {Antuono}},\ }\bibfield  {title}
  {\enquote {\bibinfo {title} {Simulating 2d open-channel flows through an sph
  model},}\ }\href@noop {} {\bibfield  {journal} {\bibinfo  {journal} {European
  Journal of Mechanics - B/Fluids}\ }\textbf {\bibinfo {volume} {34}},\
  \bibinfo {pages} {35--46} (\bibinfo {year} {2012})}\BibitemShut {NoStop}%
\bibitem [{\citenamefont {Vacondio}\ \emph {et~al.}(2012)\citenamefont
  {Vacondio}, \citenamefont {Rogers}, \citenamefont {Stansby},\ and\
  \citenamefont {Mignosa}}]{Vacondio2012mod}%
  \BibitemOpen
  \bibfield  {author} {\bibinfo {author} {\bibfnamefont {R.}~\bibnamefont
  {Vacondio}}, \bibinfo {author} {\bibfnamefont {B.}~\bibnamefont {Rogers}},
  \bibinfo {author} {\bibfnamefont {P.}~\bibnamefont {Stansby}}, \ and\
  \bibinfo {author} {\bibfnamefont {P.}~\bibnamefont {Mignosa}},\ }\bibfield
  {title} {\enquote {\bibinfo {title} {{SPH modeling of shallow flow with open
  boundaries for practical flood simulation}},}\ }\href@noop {} {\bibfield
  {journal} {\bibinfo  {journal} {Journal of Hydraulic Engineering}\ }\textbf
  {\bibinfo {volume} {138}},\ \bibinfo {pages} {530--541} (\bibinfo {year}
  {2012})}\BibitemShut {NoStop}%
\bibitem [{\citenamefont {Alvarado-Rodriguez}\ \emph
  {et~al.}(2017)\citenamefont {Alvarado-Rodriguez}, \citenamefont {Klapp},
  \citenamefont {Sigalotti}, \citenamefont {Dominguez},\ and\ \citenamefont
  {Sanchez}}]{Alvarado-Rodriguez2017}%
  \BibitemOpen
  \bibfield  {author} {\bibinfo {author} {\bibfnamefont {C.}~\bibnamefont
  {Alvarado-Rodriguez}}, \bibinfo {author} {\bibfnamefont {J.}~\bibnamefont
  {Klapp}}, \bibinfo {author} {\bibfnamefont {L.}~\bibnamefont {Sigalotti}},
  \bibinfo {author} {\bibfnamefont {J.}~\bibnamefont {Dominguez}}, \ and\
  \bibinfo {author} {\bibfnamefont {E.}~\bibnamefont {Sanchez}},\ }\bibfield
  {title} {\enquote {\bibinfo {title} {{Nonreflecting outlet boundary
  conditions for incompressible flows using SPH}},}\ }\href@noop {} {\bibfield
  {journal} {\bibinfo  {journal} {Computers and Fluids}\ }\textbf {\bibinfo
  {volume} {159}},\ \bibinfo {pages} {177--188} (\bibinfo {year}
  {2017})}\BibitemShut {NoStop}%
\bibitem [{\citenamefont {Ferrand}\ \emph {et~al.}(2017)\citenamefont
  {Ferrand}, \citenamefont {Joly}, \citenamefont {Kassiotis}, \citenamefont
  {Violeau}, \citenamefont {Leroy}, \citenamefont {Morel},\ and\ \citenamefont
  {Rogers}}]{Ferrand2017}%
  \BibitemOpen
  \bibfield  {author} {\bibinfo {author} {\bibfnamefont {M.}~\bibnamefont
  {Ferrand}}, \bibinfo {author} {\bibfnamefont {A.}~\bibnamefont {Joly}},
  \bibinfo {author} {\bibfnamefont {C.}~\bibnamefont {Kassiotis}}, \bibinfo
  {author} {\bibfnamefont {D.}~\bibnamefont {Violeau}}, \bibinfo {author}
  {\bibfnamefont {A.}~\bibnamefont {Leroy}}, \bibinfo {author} {\bibfnamefont
  {F.}~\bibnamefont {Morel}}, \ and\ \bibinfo {author} {\bibfnamefont
  {B.}~\bibnamefont {Rogers}},\ }\bibfield  {title} {\enquote {\bibinfo {title}
  {Unsteady open boundaries for {SPH} using semi-analytical conditions and
  riemann solver in {2D}},}\ }\href@noop {} {\bibfield  {journal} {\bibinfo
  {journal} {Computer Physics Communications}\ }\textbf {\bibinfo {volume}
  {210}},\ \bibinfo {pages} {29--44} (\bibinfo {year} {2017})}\BibitemShut
  {NoStop}%
\bibitem [{\citenamefont {Negi}, \citenamefont {Ramachandran},\ and\
  \citenamefont {Haftu}(2020)}]{Negi2020}%
  \BibitemOpen
  \bibfield  {author} {\bibinfo {author} {\bibfnamefont {P.}~\bibnamefont
  {Negi}}, \bibinfo {author} {\bibfnamefont {P.}~\bibnamefont {Ramachandran}},
  \ and\ \bibinfo {author} {\bibfnamefont {A.}~\bibnamefont {Haftu}},\
  }\bibfield  {title} {\enquote {\bibinfo {title} {{An improved non-reflecting
  outlet boundary condition for weakly-compressible SPH}},}\ }\href@noop {}
  {\bibfield  {journal} {\bibinfo  {journal} {Computer Methods in Applied
  Mechanics and Engineering}\ }\textbf {\bibinfo {volume} {367}},\ \bibinfo
  {pages} {113119} (\bibinfo {year} {2020})}\BibitemShut {NoStop}%
\bibitem [{\citenamefont {Martin}, \citenamefont {Basa},\ and\ \citenamefont
  {Quinlan}(2009)}]{Martin2009}%
  \BibitemOpen
  \bibfield  {author} {\bibinfo {author} {\bibfnamefont {L.}~\bibnamefont
  {Martin}}, \bibinfo {author} {\bibfnamefont {M.}~\bibnamefont {Basa}}, \ and\
  \bibinfo {author} {\bibfnamefont {N.}~\bibnamefont {Quinlan}},\ }\bibfield
  {title} {\enquote {\bibinfo {title} {Permeable and non‐reflecting boundary
  conditions in {SPH}},}\ }\href@noop {} {\bibfield  {journal} {\bibinfo
  {journal} {International Journal for Numerical Methods in Fluids}\ }\textbf
  {\bibinfo {volume} {61}},\ \bibinfo {pages} {709--724} (\bibinfo {year}
  {2009})}\BibitemShut {NoStop}%
\bibitem [{\citenamefont {Tafuni}\ \emph {et~al.}(2018)\citenamefont {Tafuni},
  \citenamefont {Domínguez}, \citenamefont {Vacondio},\ and\ \citenamefont
  {Crespo}}]{Tafuni2018}%
  \BibitemOpen
  \bibfield  {author} {\bibinfo {author} {\bibfnamefont {A.}~\bibnamefont
  {Tafuni}}, \bibinfo {author} {\bibfnamefont {J.}~\bibnamefont {Domínguez}},
  \bibinfo {author} {\bibfnamefont {R.}~\bibnamefont {Vacondio}}, \ and\
  \bibinfo {author} {\bibfnamefont {A.}~\bibnamefont {Crespo}},\ }\bibfield
  {title} {\enquote {\bibinfo {title} {A versatile algorithm for the treatment
  of open boundary conditions in smoothed particle hydrodynamics {GPU}
  models},}\ }\href@noop {} {\bibfield  {journal} {\bibinfo  {journal}
  {Computer Methods in Applied Mechanics and Engineering}\ }\textbf {\bibinfo
  {volume} {342}},\ \bibinfo {pages} {604--624} (\bibinfo {year}
  {2018})}\BibitemShut {NoStop}%
\bibitem [{\citenamefont {Zhang}\ \emph {et~al.}(2023)\citenamefont {Zhang},
  \citenamefont {Zhang}, \citenamefont {Zhang},\ and\ \citenamefont
  {Hu}}]{Shuoguo2022free}%
  \BibitemOpen
  \bibfield  {author} {\bibinfo {author} {\bibfnamefont {S.}~\bibnamefont
  {Zhang}}, \bibinfo {author} {\bibfnamefont {W.}~\bibnamefont {Zhang}},
  \bibinfo {author} {\bibfnamefont {C.}~\bibnamefont {Zhang}}, \ and\ \bibinfo
  {author} {\bibfnamefont {X.}~\bibnamefont {Hu}},\ }\bibfield  {title}
  {\enquote {\bibinfo {title} {A lagrangian free-stream boundary condition for
  weakly compressible smoothed particle hydrodynamics},}\ }\href@noop {}
  {\bibfield  {journal} {\bibinfo  {journal} {Journal of Computational
  Physics}\ }\textbf {\bibinfo {volume} {490}},\ \bibinfo {pages} {112303}
  (\bibinfo {year} {2023})}\BibitemShut {NoStop}%
\bibitem [{\citenamefont {Wang}\ \emph {et~al.}(2019)\citenamefont {Wang},
  \citenamefont {Zhang}, \citenamefont {Ming}, \citenamefont {Sun},\ and\
  \citenamefont {Cheng}}]{Wang2019}%
  \BibitemOpen
  \bibfield  {author} {\bibinfo {author} {\bibfnamefont {P.}~\bibnamefont
  {Wang}}, \bibinfo {author} {\bibfnamefont {A.}~\bibnamefont {Zhang}},
  \bibinfo {author} {\bibfnamefont {F.}~\bibnamefont {Ming}}, \bibinfo {author}
  {\bibfnamefont {P.}~\bibnamefont {Sun}}, \ and\ \bibinfo {author}
  {\bibfnamefont {H.}~\bibnamefont {Cheng}},\ }\bibfield  {title} {\enquote
  {\bibinfo {title} {{A novel nonreflecting boundary condition for fluid
  dynamics solved by smoothed particle hydrodynamics}},}\ }\href@noop {}
  {\bibfield  {journal} {\bibinfo  {journal} {Journal of Fluid Mechanics}\
  }\textbf {\bibinfo {volume} {860}},\ \bibinfo {pages} {81--114} (\bibinfo
  {year} {2019})}\BibitemShut {NoStop}%
\bibitem [{\citenamefont {Lyu}\ \emph {et~al.}(2022)\citenamefont {Lyu},
  \citenamefont {Sun}, \citenamefont {Colagrossi},\ and\ \citenamefont
  {Zhang}}]{Lyu2022Toward}%
  \BibitemOpen
  \bibfield  {author} {\bibinfo {author} {\bibfnamefont {H.-G.}\ \bibnamefont
  {Lyu}}, \bibinfo {author} {\bibfnamefont {P.}~\bibnamefont {Sun}}, \bibinfo
  {author} {\bibfnamefont {A.}~\bibnamefont {Colagrossi}}, \ and\ \bibinfo
  {author} {\bibfnamefont {A.-M.}\ \bibnamefont {Zhang}},\ }\bibfield  {title}
  {\enquote {\bibinfo {title} {Towards sph simulations of cavitating flows with
  an eosb cavitation model},}\ }\href@noop {} {\bibfield  {journal} {\bibinfo
  {journal} {Acta Mechanica Sinica}\ }\textbf {\bibinfo {volume} {39}}
  (\bibinfo {year} {2022})}\BibitemShut {NoStop}%
\bibitem [{\citenamefont {Hirschler}\ \emph {et~al.}(2016)\citenamefont
  {Hirschler}, \citenamefont {Kunz}, \citenamefont {Huber}, \citenamefont
  {Hahn},\ and\ \citenamefont {Nieken}}]{HIRSCHLER2016Open}%
  \BibitemOpen
  \bibfield  {author} {\bibinfo {author} {\bibfnamefont {M.}~\bibnamefont
  {Hirschler}}, \bibinfo {author} {\bibfnamefont {P.}~\bibnamefont {Kunz}},
  \bibinfo {author} {\bibfnamefont {M.}~\bibnamefont {Huber}}, \bibinfo
  {author} {\bibfnamefont {F.}~\bibnamefont {Hahn}}, \ and\ \bibinfo {author}
  {\bibfnamefont {U.}~\bibnamefont {Nieken}},\ }\bibfield  {title} {\enquote
  {\bibinfo {title} {Open boundary conditions for isph and their application to
  micro-flow},}\ }\href@noop {} {\bibfield  {journal} {\bibinfo  {journal}
  {Journal of Computational Physics}\ }\textbf {\bibinfo {volume} {307}},\
  \bibinfo {pages} {614--633} (\bibinfo {year} {2016})}\BibitemShut {NoStop}%
\bibitem [{\citenamefont {Kunz}\ \emph {et~al.}(2016)\citenamefont {Kunz},
  \citenamefont {Hirschler}, \citenamefont {Huber},\ and\ \citenamefont
  {Nieken}}]{KUNZ2016Inflow}%
  \BibitemOpen
  \bibfield  {author} {\bibinfo {author} {\bibfnamefont {P.}~\bibnamefont
  {Kunz}}, \bibinfo {author} {\bibfnamefont {M.}~\bibnamefont {Hirschler}},
  \bibinfo {author} {\bibfnamefont {M.}~\bibnamefont {Huber}}, \ and\ \bibinfo
  {author} {\bibfnamefont {U.}~\bibnamefont {Nieken}},\ }\bibfield  {title}
  {\enquote {\bibinfo {title} {Inflow/outflow with dirichlet boundary
  conditions for pressure in isph},}\ }\href@noop {} {\bibfield  {journal}
  {\bibinfo  {journal} {Journal of Computational Physics}\ }\textbf {\bibinfo
  {volume} {326}},\ \bibinfo {pages} {171--187} (\bibinfo {year}
  {2016})}\BibitemShut {NoStop}%
\bibitem [{\citenamefont {Monteleone}, \citenamefont {Monteforte},\ and\
  \citenamefont {Napoli}(2017)}]{Alessandra2017}%
  \BibitemOpen
  \bibfield  {author} {\bibinfo {author} {\bibfnamefont {A.}~\bibnamefont
  {Monteleone}}, \bibinfo {author} {\bibfnamefont {M.}~\bibnamefont
  {Monteforte}}, \ and\ \bibinfo {author} {\bibfnamefont {E.}~\bibnamefont
  {Napoli}},\ }\bibfield  {title} {\enquote {\bibinfo {title} {Inflow/outflow
  pressure boundary conditions for smoothed particle hydrodynamics simulations
  of incompressible flows},}\ }\href@noop {} {\bibfield  {journal} {\bibinfo
  {journal} {Computers and Fluids}\ }\textbf {\bibinfo {volume} {159}},\
  \bibinfo {pages} {9--22} (\bibinfo {year} {2017})}\BibitemShut {NoStop}%
\bibitem [{\citenamefont {Leroy}\ \emph {et~al.}(2016)\citenamefont {Leroy},
  \citenamefont {Violeau}, \citenamefont {Ferrand}, \citenamefont {Fratter},\
  and\ \citenamefont {Joly}}]{LEROY2016new}%
  \BibitemOpen
  \bibfield  {author} {\bibinfo {author} {\bibfnamefont {A.}~\bibnamefont
  {Leroy}}, \bibinfo {author} {\bibfnamefont {D.}~\bibnamefont {Violeau}},
  \bibinfo {author} {\bibfnamefont {M.}~\bibnamefont {Ferrand}}, \bibinfo
  {author} {\bibfnamefont {L.}~\bibnamefont {Fratter}}, \ and\ \bibinfo
  {author} {\bibfnamefont {A.}~\bibnamefont {Joly}},\ }\bibfield  {title}
  {\enquote {\bibinfo {title} {A new open boundary formulation for
  incompressible sph},}\ }\href@noop {} {\bibfield  {journal} {\bibinfo
  {journal} {Computers \& Mathematics with Applications}\ }\textbf {\bibinfo
  {volume} {72}},\ \bibinfo {pages} {2417--2432} (\bibinfo {year}
  {2016})}\BibitemShut {NoStop}%
\bibitem [{\citenamefont {Zhang}\ \emph
  {et~al.}(2021{\natexlab{a}})\citenamefont {Zhang}, \citenamefont {Rezavand},
  \citenamefont {Zhu}, \citenamefont {Yu}, \citenamefont {Wu}, \citenamefont
  {Zhang}, \citenamefont {Wang},\ and\ \citenamefont {Hu}}]{Zhang2021CPC}%
  \BibitemOpen
  \bibfield  {author} {\bibinfo {author} {\bibfnamefont {C.}~\bibnamefont
  {Zhang}}, \bibinfo {author} {\bibfnamefont {M.}~\bibnamefont {Rezavand}},
  \bibinfo {author} {\bibfnamefont {Y.}~\bibnamefont {Zhu}}, \bibinfo {author}
  {\bibfnamefont {Y.}~\bibnamefont {Yu}}, \bibinfo {author} {\bibfnamefont
  {D.}~\bibnamefont {Wu}}, \bibinfo {author} {\bibfnamefont {W.}~\bibnamefont
  {Zhang}}, \bibinfo {author} {\bibfnamefont {J.}~\bibnamefont {Wang}}, \ and\
  \bibinfo {author} {\bibfnamefont {X.}~\bibnamefont {Hu}},\ }\bibfield
  {title} {\enquote {\bibinfo {title} {{SPHinXsys: An open source multi-physics
  and multi-resolution library based on smoothed particle hydrodynamics}},}\
  }\href@noop {} {\bibfield  {journal} {\bibinfo  {journal} {Computer Physics
  Communications}\ }\textbf {\bibinfo {volume} {267}},\ \bibinfo {pages}
  {108066} (\bibinfo {year} {2021}{\natexlab{a}})}\BibitemShut {NoStop}%
\bibitem [{\citenamefont {Monaghan}(1994)}]{Monaghan1994Sim}%
  \BibitemOpen
  \bibfield  {author} {\bibinfo {author} {\bibfnamefont {J.}~\bibnamefont
  {Monaghan}},\ }\bibfield  {title} {\enquote {\bibinfo {title} {{Simulating
  free surface flows with SPH}},}\ }\href@noop {} {\bibfield  {journal}
  {\bibinfo  {journal} {Journal of Computational Physics}\ }\textbf {\bibinfo
  {volume} {110}},\ \bibinfo {pages} {399--406} (\bibinfo {year}
  {1994})}\BibitemShut {NoStop}%
\bibitem [{\citenamefont {Hu}\ and\ \citenamefont {Adams}(2006)}]{Hu2006multi}%
  \BibitemOpen
  \bibfield  {author} {\bibinfo {author} {\bibfnamefont {X.}~\bibnamefont
  {Hu}}\ and\ \bibinfo {author} {\bibfnamefont {N.}~\bibnamefont {Adams}},\
  }\bibfield  {title} {\enquote {\bibinfo {title} {{A multi phase SPH method
  for macroscopic and mesoscopic flows}},}\ }\href@noop {} {\bibfield
  {journal} {\bibinfo  {journal} {Journal of Computational Physics}\ }\textbf
  {\bibinfo {volume} {213}},\ \bibinfo {pages} {844--861} (\bibinfo {year}
  {2006})}\BibitemShut {NoStop}%
\bibitem [{\citenamefont {Zhang}, \citenamefont {Hu},\ and\ \citenamefont
  {Adams}(2017{\natexlab{a}})}]{Zhang2017trans}%
  \BibitemOpen
  \bibfield  {author} {\bibinfo {author} {\bibfnamefont {C.}~\bibnamefont
  {Zhang}}, \bibinfo {author} {\bibfnamefont {X.}~\bibnamefont {Hu}}, \ and\
  \bibinfo {author} {\bibfnamefont {N.}~\bibnamefont {Adams}},\ }\bibfield
  {title} {\enquote {\bibinfo {title} {A generalized transport-velocity
  formulation for smoothed particle hydrodynamics},}\ }\href@noop {} {\bibfield
   {journal} {\bibinfo  {journal} {Journal of Computational Physics}\ }\textbf
  {\bibinfo {volume} {337}},\ \bibinfo {pages} {216--232} (\bibinfo {year}
  {2017}{\natexlab{a}})}\BibitemShut {NoStop}%
\bibitem [{\citenamefont {Zhang}, \citenamefont {Rezavand},\ and\ \citenamefont
  {Hu}(2020)}]{Zhang2020Dual}%
  \BibitemOpen
  \bibfield  {author} {\bibinfo {author} {\bibfnamefont {C.}~\bibnamefont
  {Zhang}}, \bibinfo {author} {\bibfnamefont {M.}~\bibnamefont {Rezavand}}, \
  and\ \bibinfo {author} {\bibfnamefont {X.}~\bibnamefont {Hu}},\ }\bibfield
  {title} {\enquote {\bibinfo {title} {{Dual-criteria time stepping for weakly
  compressible smoothed particle hydrodynamics}},}\ }\href@noop {} {\bibfield
  {journal} {\bibinfo  {journal} {Journal of Computational Physics}\ }\textbf
  {\bibinfo {volume} {404}},\ \bibinfo {pages} {109135} (\bibinfo {year}
  {2020})}\BibitemShut {NoStop}%
\bibitem [{\citenamefont {Zhang}\ \emph
  {et~al.}(2021{\natexlab{b}})\citenamefont {Zhang}, \citenamefont {Wei},
  \citenamefont {Dias},\ and\ \citenamefont {Hu}}]{ZHANG2021An}%
  \BibitemOpen
  \bibfield  {author} {\bibinfo {author} {\bibfnamefont {C.}~\bibnamefont
  {Zhang}}, \bibinfo {author} {\bibfnamefont {Y.}~\bibnamefont {Wei}}, \bibinfo
  {author} {\bibfnamefont {F.}~\bibnamefont {Dias}}, \ and\ \bibinfo {author}
  {\bibfnamefont {X.}~\bibnamefont {Hu}},\ }\bibfield  {title} {\enquote
  {\bibinfo {title} {An efficient fully lagrangian solver for modeling wave
  interaction with oscillating wave surge converter},}\ }\href@noop {}
  {\bibfield  {journal} {\bibinfo  {journal} {Ocean Engineering}\ }\textbf
  {\bibinfo {volume} {236}},\ \bibinfo {pages} {109540} (\bibinfo {year}
  {2021}{\natexlab{b}})}\BibitemShut {NoStop}%
\bibitem [{\citenamefont {Zhang}, \citenamefont {Hu},\ and\ \citenamefont
  {Adams}(2017{\natexlab{b}})}]{article2017Chi}%
  \BibitemOpen
  \bibfield  {author} {\bibinfo {author} {\bibfnamefont {C.}~\bibnamefont
  {Zhang}}, \bibinfo {author} {\bibfnamefont {X.}~\bibnamefont {Hu}}, \ and\
  \bibinfo {author} {\bibfnamefont {N.}~\bibnamefont {Adams}},\ }\bibfield
  {title} {\enquote {\bibinfo {title} {A weakly compressible {SPH} method based
  on a low-dissipation {R}iemann solver},}\ }\href@noop {} {\bibfield
  {journal} {\bibinfo  {journal} {Journal of Computational Physics}\ }\textbf
  {\bibinfo {volume} {335}} (\bibinfo {year} {2017}{\natexlab{b}})}\BibitemShut
  {NoStop}%
\bibitem [{\citenamefont {Lind}\ \emph {et~al.}(2012)\citenamefont {Lind},
  \citenamefont {Xu}, \citenamefont {Stansby},\ and\ \citenamefont
  {Rogers}}]{Lind2012Incom}%
  \BibitemOpen
  \bibfield  {author} {\bibinfo {author} {\bibfnamefont {S.}~\bibnamefont
  {Lind}}, \bibinfo {author} {\bibfnamefont {R.}~\bibnamefont {Xu}}, \bibinfo
  {author} {\bibfnamefont {P.}~\bibnamefont {Stansby}}, \ and\ \bibinfo
  {author} {\bibfnamefont {B.}~\bibnamefont {Rogers}},\ }\bibfield  {title}
  {\enquote {\bibinfo {title} {Incompressible smoothed particle hydrodynamics
  for free-surface flows: A generalised diffusion-based algorithm for stability
  and validations for impulsive flows and propagating waves.}}\ }\href@noop {}
  {\bibfield  {journal} {\bibinfo  {journal} {Journal of Computational
  Physics}\ }\textbf {\bibinfo {volume} {231}},\ \bibinfo {pages} {1499--1523}
  (\bibinfo {year} {2012})}\BibitemShut {NoStop}%
\bibitem [{\citenamefont {Oger}\ \emph {et~al.}(2016)\citenamefont {Oger},
  \citenamefont {Marrone}, \citenamefont {{Le Touzé}},\ and\ \citenamefont
  {{de Leffe}}}]{OGER201676SPH}%
  \BibitemOpen
  \bibfield  {author} {\bibinfo {author} {\bibfnamefont {G.}~\bibnamefont
  {Oger}}, \bibinfo {author} {\bibfnamefont {S.}~\bibnamefont {Marrone}},
  \bibinfo {author} {\bibfnamefont {D.}~\bibnamefont {{Le Touzé}}}, \ and\
  \bibinfo {author} {\bibfnamefont {M.}~\bibnamefont {{de Leffe}}},\ }\bibfield
   {title} {\enquote {\bibinfo {title} {Sph accuracy improvement through the
  combination of a quasi-lagrangian shifting transport velocity and consistent
  ale formalisms},}\ }\href@noop {} {\bibfield  {journal} {\bibinfo  {journal}
  {Journal of Computational Physics}\ }\textbf {\bibinfo {volume} {313}},\
  \bibinfo {pages} {76--98} (\bibinfo {year} {2016})}\BibitemShut {NoStop}%
\bibitem [{\citenamefont {Litvinov}, \citenamefont {Hu},\ and\ \citenamefont
  {Adams}(2015)}]{litvinov2015towards}%
  \BibitemOpen
  \bibfield  {author} {\bibinfo {author} {\bibfnamefont {S.}~\bibnamefont
  {Litvinov}}, \bibinfo {author} {\bibfnamefont {X.}~\bibnamefont {Hu}}, \ and\
  \bibinfo {author} {\bibfnamefont {N.~A.}\ \bibnamefont {Adams}},\ }\bibfield
  {title} {\enquote {\bibinfo {title} {Towards consistence and convergence of
  conservative sph approximations},}\ }\href@noop {} {\bibfield  {journal}
  {\bibinfo  {journal} {Journal of Computational Physics}\ }\textbf {\bibinfo
  {volume} {301}},\ \bibinfo {pages} {394--401} (\bibinfo {year}
  {2015})}\BibitemShut {NoStop}%
\bibitem [{\citenamefont {Negi}\ and\ \citenamefont
  {Ramachandran}(2022{\natexlab{b}})}]{Negi2022Tec}%
  \BibitemOpen
  \bibfield  {author} {\bibinfo {author} {\bibfnamefont {P.}~\bibnamefont
  {Negi}}\ and\ \bibinfo {author} {\bibfnamefont {P.}~\bibnamefont
  {Ramachandran}},\ }\bibfield  {title} {\enquote {\bibinfo {title} {Techniques
  for second-order convergent weakly compressible smoothed particle
  hydrodynamics schemes without boundaries},}\ }\href@noop {} {\bibfield
  {journal} {\bibinfo  {journal} {Physics of Fluids}\ }\textbf {\bibinfo
  {volume} {34}},\ \bibinfo {pages} {087125} (\bibinfo {year}
  {2022}{\natexlab{b}})}\BibitemShut {NoStop}%
\bibitem [{\citenamefont {Skillen}\ \emph {et~al.}(2013)\citenamefont
  {Skillen}, \citenamefont {Lind}, \citenamefont {Stansby},\ and\ \citenamefont
  {Rogers}}]{Skillen2013}%
  \BibitemOpen
  \bibfield  {author} {\bibinfo {author} {\bibfnamefont {A.}~\bibnamefont
  {Skillen}}, \bibinfo {author} {\bibfnamefont {S.}~\bibnamefont {Lind}},
  \bibinfo {author} {\bibfnamefont {P.}~\bibnamefont {Stansby}}, \ and\
  \bibinfo {author} {\bibfnamefont {B.}~\bibnamefont {Rogers}},\ }\bibfield
  {title} {\enquote {\bibinfo {title} {Incompressible smoothed particle
  hydrodynamics (sph) with reduced temporal noise and generalised fickian
  smoothing applied to body-water slam and efficient wave-body interaction},}\
  }\href@noop {} {\bibfield  {journal} {\bibinfo  {journal} {Computer Methods
  in Applied Mechanics and Engineering}\ }\textbf {\bibinfo {volume} {265}},\
  \bibinfo {pages} {163--173} (\bibinfo {year} {2013})}\BibitemShut {NoStop}%
\bibitem [{\citenamefont {Khayyer}, \citenamefont {Gotoh},\ and\ \citenamefont
  {Shimizu}(2017)}]{KHAYYER2017236Comp}%
  \BibitemOpen
  \bibfield  {author} {\bibinfo {author} {\bibfnamefont {A.}~\bibnamefont
  {Khayyer}}, \bibinfo {author} {\bibfnamefont {H.}~\bibnamefont {Gotoh}}, \
  and\ \bibinfo {author} {\bibfnamefont {Y.}~\bibnamefont {Shimizu}},\
  }\bibfield  {title} {\enquote {\bibinfo {title} {Comparative study on
  accuracy and conservation properties of two particle regularization schemes
  and proposal of an optimized particle shifting scheme in isph context},}\
  }\href@noop {} {\bibfield  {journal} {\bibinfo  {journal} {Journal of
  Computational Physics}\ }\textbf {\bibinfo {volume} {332}},\ \bibinfo {pages}
  {236--256} (\bibinfo {year} {2017})}\BibitemShut {NoStop}%
\bibitem [{\citenamefont {Sun}\ \emph {et~al.}(2017)\citenamefont {Sun},
  \citenamefont {Colagrossi}, \citenamefont {Marrone},\ and\ \citenamefont
  {Zhang}}]{SUN201725}%
  \BibitemOpen
  \bibfield  {author} {\bibinfo {author} {\bibfnamefont {P.}~\bibnamefont
  {Sun}}, \bibinfo {author} {\bibfnamefont {A.}~\bibnamefont {Colagrossi}},
  \bibinfo {author} {\bibfnamefont {S.}~\bibnamefont {Marrone}}, \ and\
  \bibinfo {author} {\bibfnamefont {A.}~\bibnamefont {Zhang}},\ }\bibfield
  {title} {\enquote {\bibinfo {title} {The delta plus-sph model: Simple
  procedures for a further improvement of the sph scheme},}\ }\href@noop {}
  {\bibfield  {journal} {\bibinfo  {journal} {Computer Methods in Applied
  Mechanics and Engineering}\ }\textbf {\bibinfo {volume} {315}},\ \bibinfo
  {pages} {25--49} (\bibinfo {year} {2017})}\BibitemShut {NoStop}%
\bibitem [{\citenamefont {Adami}, \citenamefont {Hu},\ and\ \citenamefont
  {Adams}(2013)}]{Adami2013trans}%
  \BibitemOpen
  \bibfield  {author} {\bibinfo {author} {\bibfnamefont {S.}~\bibnamefont
  {Adami}}, \bibinfo {author} {\bibfnamefont {X.}~\bibnamefont {Hu}}, \ and\
  \bibinfo {author} {\bibfnamefont {N.}~\bibnamefont {Adams}},\ }\bibfield
  {title} {\enquote {\bibinfo {title} {A transport-velocity formulation for
  smoothed particle hydrodynamics},}\ }\href@noop {} {\bibfield  {journal}
  {\bibinfo  {journal} {Journal of Computational Physics}\ }\textbf {\bibinfo
  {volume} {241}},\ \bibinfo {pages} {292--307} (\bibinfo {year}
  {2013})}\BibitemShut {NoStop}%
\bibitem [{\citenamefont {Dilts}(1999)}]{Dilts}%
  \BibitemOpen
  \bibfield  {author} {\bibinfo {author} {\bibfnamefont {G.}~\bibnamefont
  {Dilts}},\ }\bibfield  {title} {\enquote {\bibinfo {title}
  {{Moving-least-squares-particle hydrodynamics—I. Consistency and
  stability}},}\ }\href@noop {} {\bibfield  {journal} {\bibinfo  {journal}
  {International Journal for Numerical Methods in Engineering}\ }\textbf
  {\bibinfo {volume} {44}},\ \bibinfo {pages} {1115--1155} (\bibinfo {year}
  {1999})}\BibitemShut {NoStop}%
\bibitem [{\citenamefont {Haque}\ and\ \citenamefont {Dilts}(2007)}]{Haque}%
  \BibitemOpen
  \bibfield  {author} {\bibinfo {author} {\bibfnamefont {A.}~\bibnamefont
  {Haque}}\ and\ \bibinfo {author} {\bibfnamefont {G.}~\bibnamefont {Dilts}},\
  }\bibfield  {title} {\enquote {\bibinfo {title} {Three-dimensional boundary
  detection for particle methods},}\ }\href@noop {} {\bibfield  {journal}
  {\bibinfo  {journal} {Journal of Computational Physics}\ }\textbf {\bibinfo
  {volume} {226}},\ \bibinfo {pages} {1710--1730} (\bibinfo {year}
  {2007})}\BibitemShut {NoStop}%
\bibitem [{\citenamefont {Lee}\ \emph {et~al.}(2008)\citenamefont {Lee},
  \citenamefont {Moulinec}, \citenamefont {Xu}, \citenamefont {Violeau},
  \citenamefont {Laurence},\ and\ \citenamefont {Stansby}}]{Lee2008}%
  \BibitemOpen
  \bibfield  {author} {\bibinfo {author} {\bibfnamefont {E.}~\bibnamefont
  {Lee}}, \bibinfo {author} {\bibfnamefont {C.}~\bibnamefont {Moulinec}},
  \bibinfo {author} {\bibfnamefont {R.}~\bibnamefont {Xu}}, \bibinfo {author}
  {\bibfnamefont {D.}~\bibnamefont {Violeau}}, \bibinfo {author} {\bibfnamefont
  {D.}~\bibnamefont {Laurence}}, \ and\ \bibinfo {author} {\bibfnamefont
  {P.}~\bibnamefont {Stansby}},\ }\bibfield  {title} {\enquote {\bibinfo
  {title} {Comparisons of weakly compressible and truly incompressible
  algorithms for the {SPH} mesh free particle method},}\ }\href@noop {}
  {\bibfield  {journal} {\bibinfo  {journal} {Journal of Computational
  Physics}\ }\textbf {\bibinfo {volume} {227}},\ \bibinfo {pages} {8417--8436}
  (\bibinfo {year} {2008})}\BibitemShut {NoStop}%
\bibitem [{\citenamefont {Wendland}(1995)}]{Wendland1995}%
  \BibitemOpen
  \bibfield  {author} {\bibinfo {author} {\bibfnamefont {H.}~\bibnamefont
  {Wendland}},\ }\bibfield  {title} {\enquote {\bibinfo {title} {{Piecewise
  polynomial, positive definite and compactly supported radial functions of
  minimal degree}},}\ }\href@noop {} {\bibfield  {journal} {\bibinfo  {journal}
  {Advances in Computational Mathematics}\ }\textbf {\bibinfo {volume} {4}},\
  \bibinfo {pages} {389--396} (\bibinfo {year} {1995})}\BibitemShut {NoStop}%
\bibitem [{\citenamefont {Cheng}\ and\ \citenamefont
  {Gupta}(1989)}]{Cheng1989an}%
  \BibitemOpen
  \bibfield  {author} {\bibinfo {author} {\bibfnamefont {H.}~\bibnamefont
  {Cheng}}\ and\ \bibinfo {author} {\bibfnamefont {K.~C.}\ \bibnamefont
  {Gupta}},\ }\bibfield  {title} {\enquote {\bibinfo {title} {{An Historical
  Note on Finite Rotations}},}\ }\href@noop {} {\bibfield  {journal} {\bibinfo
  {journal} {Journal of Applied Mechanics}\ }\textbf {\bibinfo {volume} {56}},\
  \bibinfo {pages} {139--145} (\bibinfo {year} {1989})}\BibitemShut {NoStop}%
\bibitem [{\citenamefont {Fraiture}(2009)}]{Fraiture2009his}%
  \BibitemOpen
  \bibfield  {author} {\bibinfo {author} {\bibfnamefont {L.}~\bibnamefont
  {Fraiture}},\ }\bibfield  {title} {\enquote {\bibinfo {title} {A history of
  the description of the three-dimensional finite rotation},}\ }\href@noop {}
  {\bibfield  {journal} {\bibinfo  {journal} {The Journal of the Astronautical
  Sciences}\ }\textbf {\bibinfo {volume} {57}},\ \bibinfo {pages} {207--232}
  (\bibinfo {year} {2009})}\BibitemShut {NoStop}%
\bibitem [{\citenamefont {Dai}(2015)}]{DAI2015Euler}%
  \BibitemOpen
  \bibfield  {author} {\bibinfo {author} {\bibfnamefont {J.~S.}\ \bibnamefont
  {Dai}},\ }\bibfield  {title} {\enquote {\bibinfo {title} {Euler–rodrigues
  formula variations, quaternion conjugation and intrinsic connections},}\
  }\href@noop {} {\bibfield  {journal} {\bibinfo  {journal} {Mechanism and
  Machine Theory}\ }\textbf {\bibinfo {volume} {92}},\ \bibinfo {pages}
  {144--152} (\bibinfo {year} {2015})}\BibitemShut {NoStop}%
\bibitem [{\citenamefont {Zhang}\ \emph {et~al.}(2024)\citenamefont {Zhang},
  \citenamefont {Fan}, \citenamefont {Wu}, \citenamefont {Zhang},\ and\
  \citenamefont {Hu}}]{zhang2024dynamical}%
  \BibitemOpen
  \bibfield  {author} {\bibinfo {author} {\bibfnamefont {S.}~\bibnamefont
  {Zhang}}, \bibinfo {author} {\bibfnamefont {Y.}~\bibnamefont {Fan}}, \bibinfo
  {author} {\bibfnamefont {D.}~\bibnamefont {Wu}}, \bibinfo {author}
  {\bibfnamefont {C.}~\bibnamefont {Zhang}}, \ and\ \bibinfo {author}
  {\bibfnamefont {X.}~\bibnamefont {Hu}},\ }\bibfield  {title} {\enquote
  {\bibinfo {title} {Dynamical pressure boundary condition for
  weakly-compressible smoothed particle hydrodynamics},}\ }\href@noop {}
  {\bibfield  {journal} {\bibinfo  {journal} {arXiv preprint arXiv:2403.09485}\
  } (\bibinfo {year} {2024})}\BibitemShut {NoStop}%
\bibitem [{\citenamefont {{Morris}}, \citenamefont {{Fox}},\ and\ \citenamefont
  {{Zhu}}(1997)}]{Morris1997Modeling}%
  \BibitemOpen
  \bibfield  {author} {\bibinfo {author} {\bibfnamefont {J.~P.}\ \bibnamefont
  {{Morris}}}, \bibinfo {author} {\bibfnamefont {P.~J.}\ \bibnamefont {{Fox}}},
  \ and\ \bibinfo {author} {\bibfnamefont {Y.}~\bibnamefont {{Zhu}}},\
  }\bibfield  {title} {\enquote {\bibinfo {title} {{Modeling Low Reynolds
  Number Incompressible Flows Using SPH}},}\ }\href@noop {} {\bibfield
  {journal} {\bibinfo  {journal} {Journal of Computational Physics}\ }\textbf
  {\bibinfo {volume} {136}},\ \bibinfo {pages} {214--226} (\bibinfo {year}
  {1997})}\BibitemShut {NoStop}%
\bibitem [{\citenamefont {{Takeda}}, \citenamefont {{Miyama}},\ and\
  \citenamefont {{Sekiya}}(1994)}]{Takeda1994Nume}%
  \BibitemOpen
  \bibfield  {author} {\bibinfo {author} {\bibfnamefont {H.}~\bibnamefont
  {{Takeda}}}, \bibinfo {author} {\bibfnamefont {S.~M.}\ \bibnamefont
  {{Miyama}}}, \ and\ \bibinfo {author} {\bibfnamefont {M.}~\bibnamefont
  {{Sekiya}}},\ }\bibfield  {title} {\enquote {\bibinfo {title} {{Numerical
  Simulation of Viscous Flow by Smoothed Particle Hydrodynamics}},}\
  }\href@noop {} {\bibfield  {journal} {\bibinfo  {journal} {Progress of
  Theoretical Physics}\ }\textbf {\bibinfo {volume} {92}},\ \bibinfo {pages}
  {939--960} (\bibinfo {year} {1994})}\BibitemShut {NoStop}%
\bibitem [{\citenamefont {{Sigalotti}}\ \emph {et~al.}(2003)\citenamefont
  {{Sigalotti}}, \citenamefont {{Klapp}}, \citenamefont {{Sira}}, \citenamefont
  {{Mele{\'a}n}},\ and\ \citenamefont {{Hasmy}}}]{Sigalotti2003SPH}%
  \BibitemOpen
  \bibfield  {author} {\bibinfo {author} {\bibfnamefont {L.~D.~G.}\
  \bibnamefont {{Sigalotti}}}, \bibinfo {author} {\bibfnamefont
  {J.}~\bibnamefont {{Klapp}}}, \bibinfo {author} {\bibfnamefont
  {E.}~\bibnamefont {{Sira}}}, \bibinfo {author} {\bibfnamefont
  {Y.}~\bibnamefont {{Mele{\'a}n}}}, \ and\ \bibinfo {author} {\bibfnamefont
  {A.}~\bibnamefont {{Hasmy}}},\ }\bibfield  {title} {\enquote {\bibinfo
  {title} {{SPH simulations of time-dependent Poiseuille flow at low Reynolds
  numbers}},}\ }\href@noop {} {\bibfield  {journal} {\bibinfo  {journal}
  {Journal of Computational Physics}\ }\textbf {\bibinfo {volume} {191}},\
  \bibinfo {pages} {622--638} (\bibinfo {year} {2003})}\BibitemShut {NoStop}%
\bibitem [{\citenamefont {Holmes}, \citenamefont {Williams},\ and\
  \citenamefont {Tilke}(2011)}]{Holmes2011Smoo}%
  \BibitemOpen
  \bibfield  {author} {\bibinfo {author} {\bibfnamefont {D.}~\bibnamefont
  {Holmes}}, \bibinfo {author} {\bibfnamefont {J.}~\bibnamefont {Williams}}, \
  and\ \bibinfo {author} {\bibfnamefont {P.}~\bibnamefont {Tilke}},\ }\bibfield
   {title} {\enquote {\bibinfo {title} {Smooth particle hydrodynamics
  simulations of low reynolds number flows through porous media},}\ }\href@noop
  {} {\bibfield  {journal} {\bibinfo  {journal} {International Journal for
  Numerical and Analytical Methods in Geomechanics}\ }\textbf {\bibinfo
  {volume} {35}},\ \bibinfo {pages} {419 -- 437} (\bibinfo {year}
  {2011})}\BibitemShut {NoStop}%
\bibitem [{\citenamefont {Adami}, \citenamefont {Hu},\ and\ \citenamefont
  {Adams}(2012)}]{ADAMI2012gene}%
  \BibitemOpen
  \bibfield  {author} {\bibinfo {author} {\bibfnamefont {S.}~\bibnamefont
  {Adami}}, \bibinfo {author} {\bibfnamefont {X.}~\bibnamefont {Hu}}, \ and\
  \bibinfo {author} {\bibfnamefont {N.}~\bibnamefont {Adams}},\ }\bibfield
  {title} {\enquote {\bibinfo {title} {A generalized wall boundary condition
  for smoothed particle hydrodynamics},}\ }\href@noop {} {\bibfield  {journal}
  {\bibinfo  {journal} {Journal of Computational Physics}\ }\textbf {\bibinfo
  {volume} {231}},\ \bibinfo {pages} {7057--7075} (\bibinfo {year}
  {2012})}\BibitemShut {NoStop}%
\bibitem [{\citenamefont {Womersley}(1955)}]{WOMERSLEY1955met}%
  \BibitemOpen
  \bibfield  {author} {\bibinfo {author} {\bibfnamefont {J.~R.}\ \bibnamefont
  {Womersley}},\ }\bibfield  {title} {\enquote {\bibinfo {title} {Method for
  the calculation of velocity, rate of flow and viscous drag in arteries when
  the pressure gradient is known},}\ }\href@noop {} {\bibfield  {journal}
  {\bibinfo  {journal} {The Journal of Physiology}\ }\textbf {\bibinfo {volume}
  {127}},\ \bibinfo {pages} {553--563} (\bibinfo {year} {1955})}\BibitemShut
  {NoStop}%
\bibitem [{\citenamefont {Laganà}\ \emph {et~al.}(2005)\citenamefont
  {Laganà}, \citenamefont {Balossino}, \citenamefont {Migliavacca},
  \citenamefont {Pennati}, \citenamefont {Bove}, \citenamefont {{de Leval}},\
  and\ \citenamefont {Dubini}}]{Katia2005Multi}%
  \BibitemOpen
  \bibfield  {author} {\bibinfo {author} {\bibfnamefont {K.}~\bibnamefont
  {Laganà}}, \bibinfo {author} {\bibfnamefont {R.}~\bibnamefont {Balossino}},
  \bibinfo {author} {\bibfnamefont {F.}~\bibnamefont {Migliavacca}}, \bibinfo
  {author} {\bibfnamefont {G.}~\bibnamefont {Pennati}}, \bibinfo {author}
  {\bibfnamefont {E.~L.}\ \bibnamefont {Bove}}, \bibinfo {author}
  {\bibfnamefont {M.~R.}\ \bibnamefont {{de Leval}}}, \ and\ \bibinfo {author}
  {\bibfnamefont {G.}~\bibnamefont {Dubini}},\ }\bibfield  {title} {\enquote
  {\bibinfo {title} {Multiscale modeling of the cardiovascular system:
  application to the study of pulmonary and coronary perfusions in the
  univentricular circulation},}\ }\href@noop {} {\bibfield  {journal} {\bibinfo
   {journal} {Journal of Biomechanics}\ }\textbf {\bibinfo {volume} {38}},\
  \bibinfo {pages} {1129--1141} (\bibinfo {year} {2005})}\BibitemShut {NoStop}%
\bibitem [{\citenamefont {Gallo}\ \emph {et~al.}(2012)\citenamefont {Gallo},
  \citenamefont {De~Santis}, \citenamefont {Negri}, \citenamefont {Tresoldi},
  \citenamefont {Ponzini}, \citenamefont {Massai}, \citenamefont {Deriu},
  \citenamefont {Segers}, \citenamefont {Verhegghe}, \citenamefont {Rizzo},\
  and\ \citenamefont {Morbiducci}}]{Gallo2012Vivo}%
  \BibitemOpen
  \bibfield  {author} {\bibinfo {author} {\bibfnamefont {D.}~\bibnamefont
  {Gallo}}, \bibinfo {author} {\bibfnamefont {G.}~\bibnamefont {De~Santis}},
  \bibinfo {author} {\bibfnamefont {F.}~\bibnamefont {Negri}}, \bibinfo
  {author} {\bibfnamefont {D.}~\bibnamefont {Tresoldi}}, \bibinfo {author}
  {\bibfnamefont {R.}~\bibnamefont {Ponzini}}, \bibinfo {author} {\bibfnamefont
  {D.}~\bibnamefont {Massai}}, \bibinfo {author} {\bibfnamefont
  {M.}~\bibnamefont {Deriu}}, \bibinfo {author} {\bibfnamefont
  {P.}~\bibnamefont {Segers}}, \bibinfo {author} {\bibfnamefont
  {B.}~\bibnamefont {Verhegghe}}, \bibinfo {author} {\bibfnamefont
  {G.}~\bibnamefont {Rizzo}}, \ and\ \bibinfo {author} {\bibfnamefont
  {U.}~\bibnamefont {Morbiducci}},\ }\bibfield  {title} {\enquote {\bibinfo
  {title} {On the use of in vivo measured flow rates as boundary conditions for
  image-based hemodynamic models of the human aorta: Implications for
  indicators of abnormal flow},}\ }\href@noop {} {\bibfield  {journal}
  {\bibinfo  {journal} {Annals of biomedical engineering}\ }\textbf {\bibinfo
  {volume} {40}},\ \bibinfo {pages} {729--41} (\bibinfo {year}
  {2012})}\BibitemShut {NoStop}%
\bibitem [{\citenamefont {Taib}\ \emph {et~al.}(2019)\citenamefont {Taib},
  \citenamefont {Paisal}, \citenamefont {Ismail}, \citenamefont
  {Tajul~Arifin},\ and\ \citenamefont {Darlis}}]{Taib2019blood}%
  \BibitemOpen
  \bibfield  {author} {\bibinfo {author} {\bibfnamefont {I.}~\bibnamefont
  {Taib}}, \bibinfo {author} {\bibfnamefont {M.~S.~A.}\ \bibnamefont {Paisal}},
  \bibinfo {author} {\bibfnamefont {A.~E.}\ \bibnamefont {Ismail}}, \bibinfo
  {author} {\bibfnamefont {A.~M.}\ \bibnamefont {Tajul~Arifin}}, \ and\
  \bibinfo {author} {\bibfnamefont {N.}~\bibnamefont {Darlis}},\ }\bibfield
  {title} {\enquote {\bibinfo {title} {An analysis of blood pressure waveform
  using windkessel model for normotensive and hypertensive conditions in
  carotid artery},}\ }\href@noop {} {\bibfield  {journal} {\bibinfo  {journal}
  {Journal of Advanced Research in Fluid Mechanics and Thermal Sciences}\
  }\textbf {\bibinfo {volume} {57}},\ \bibinfo {pages} {69--85} (\bibinfo
  {year} {2019})}\BibitemShut {NoStop}%
\bibitem [{\citenamefont {de~Almeida Jansen~Catanho}, \citenamefont {Sinha},\
  and\ \citenamefont {Vijayan}(2012)}]{Catanho2012Model}%
  \BibitemOpen
  \bibfield  {author} {\bibinfo {author} {\bibfnamefont {M.~T.}\ \bibnamefont
  {de~Almeida Jansen~Catanho}}, \bibinfo {author} {\bibfnamefont
  {M.}~\bibnamefont {Sinha}}, \ and\ \bibinfo {author} {\bibfnamefont
  {V.}~\bibnamefont {Vijayan}},\ }\bibfield  {title} {\enquote {\bibinfo
  {title} {Model of aortic blood flow using the windkessel effect},}\ \
  }(\bibinfo {year} {2012})\BibitemShut {NoStop}%
\bibitem [{Kim(2007)}]{Kim2007coupling}%
  \BibitemOpen
  \href@noop {} {\emph {\bibinfo {title} {On coupling a lumped-parameter heart
  model with a three-dimensional finite element model of the aorta}}},\ ASME
  Summer Bioengineering Conference\ (\bibinfo {year} {2007})\BibitemShut
  {NoStop}%
\bibitem [{\citenamefont {Sudharsan}\ and\ \citenamefont
  {Cherry~Kemmerling}(2018)}]{Sudharsan2018effect}%
  \BibitemOpen
  \bibfield  {author} {\bibinfo {author} {\bibfnamefont {M.}~\bibnamefont
  {Sudharsan}}\ and\ \bibinfo {author} {\bibfnamefont {E.}~\bibnamefont
  {Cherry~Kemmerling}},\ }\bibfield  {title} {\enquote {\bibinfo {title} {The
  effect of inlet and outlet boundary conditions in image-based cfd modeling of
  aortic flow},}\ }\href@noop {} {\bibfield  {journal} {\bibinfo  {journal}
  {BioMedical Engineering OnLine}\ }\textbf {\bibinfo {volume} {17}} (\bibinfo
  {year} {2018})}\BibitemShut {NoStop}%
\end{thebibliography}%
		
	\end{document}